\newif\ifsubmode
\submodefalse

\newif\ifprintfig
\printfigtrue

\ifsubmode
  \documentstyle[12pt,aasms4,epsf]{article}
  \received{}
  \accepted{}
  \journalid{}{}
  \articleid{}{}
\else
  \documentstyle[11pt,aaspp4,epsf]{article}
  \slugcomment{{\it submitted to The Astronomical Journal on June 27, 1998}}
\fi

\lefthead{van der Marel}
\righthead{The black hole mass distribution in early-type galaxies}

\newcommand{\etal}{{et al.~}}

\newcommand{\lta}{\lesssim}
\newcommand{\gta}{\gtrsim}

\newcommand{\kms}{\>{\rm km}\,{\rm s}^{-1}}
\newcommand{\pc}{\>{\rm pc}}
\newcommand{\Mpc}{\>{\rm Mpc}}
\newcommand{\Msun}{\>{\rm M_{\odot}}}
\newcommand{\Lsun}{\>{\rm L_{\odot}}}

\newcommand{\Mbh}{M_{\bullet}}
\newcommand{\rbh}{r_{\bullet}}
\newcommand{\MV}{M_V}

\newcommand{\rb}{r_{\rm b}}
\newcommand{\Ib}{I_{\rm b}}

\newcommand{\VV}{$V$}

\newcommand{\II}{$I$}

\begin{document}

\title{The black hole mass distribution in early-type galaxies:\\
cusps in HST photometry interpreted through adiabatic black hole growth}

\author{Roeland P.~van der Marel\altaffilmark{1}}
\affil{Space Telescope Science Institute, 3700 San Martin Drive, 
       Baltimore, MD 21218}


\altaffiltext{1}{STScI Fellow.}


\ifsubmode\else
\clearpage\fi


\ifsubmode\else
\baselineskip=14pt
\fi


\begin{abstract}
HST observations show that the surface brightness profiles of
early-type galaxies have central cusps, $I \propto r^{-\gamma}$. Two
characteristic profile types are observed: `core' profiles have a
break at a resolved radius and $\gamma \leq 0.3$ inside that radius;
`power-law' profiles have no clear break and $\gamma > 0.3$. With few
exceptions, galaxies with $\MV < -22$ have core profiles, and galaxies
with $\MV > -20.5$ have power-law profiles. Both profile types occur
in galaxies with $-22 < \MV < -20.5$. We show here that these results
are consistent with the hypothesis that: (i) all early-type galaxies
have central BHs that grew adiabatically in homogeneous isothermal
cores; and (ii) these `progenitor' cores followed scaling relations
similar to those of the fundamental plane.

The models studied here are the ones first proposed by Young. They
predict $I \propto r^{-1/2}$ at asymptotically small radii, but $I
\propto r^{-\gamma}$ at the radii observable with HST. The slope
$\gamma$ can take all observed values; it increases monotonically with
$\mu = \Mbh / M_{\rm core}$. The scaling relations for early-type
galaxies imply that the progenitor core mass scales with luminosity as
$M_{\rm core} \propto L^{1.5}$. If, as suggested by various arguments,
the black hole (BH) mass $\Mbh$ scales roughly linearly with
luminosity, $\Mbh \propto L$, then $\mu \propto L^{-0.5}$. This yields
larger cusp slopes in lower-luminosity galaxies. Models with BH masses
and progenitor cores that obey established scaling relations predict
(at the distance of the Virgo cluster) that galaxies with $\MV <
-21.2$ have core profiles and galaxies with $\MV > -21.2$ have
power-law profiles. This reproduces both the sense and the absolute
magnitude of the observed transition. Intrinsic scatter in BH and
galaxy properties can explain why both types of galaxies are observed
around the transition magnitude. The observed bimodality in cusp
slopes may be due to a bimodality in $\Mbh/L$, with rapidly rotating
disky galaxies having larger $\Mbh/L$ than slowly rotating boxy
galaxies.

We apply the models to 46 galaxies with published HST photometry. Both
core and power-law galaxies are well fitted. The models suggest a
roughly linear correlation between BH mass and {\VV}-band galaxy
luminosity, $\log \Mbh \approx -1.83 + \log L$ in solar units (RMS
scatter $0.33$ dex). This agrees with the average relation for nearby
galaxies with kinematically determined BH masses. Photometrically and
kinematically determined BH masses agree to within $\sim 0.25$ dex RMS
for galaxies that have both. These results provide additional support
to the hypothesis that every galaxy (spheroid) has a central BH. The
BH mass distribution inferred here is consistent with quasar
statistics for a BH accretion efficiency of $4$\%.

The proposed scenario is not a unique way to interpret the observed
surface brightness cusps of galaxies, but it explains observational
correlations that are otherwise unexplained, and it yields BH masses
that agree with those determined kinematically.
\end{abstract}



\clearpage


\section{Introduction}
\label{s:intro}

The high spatial resolution of the Hubble Space Telescope (HST) has
allowed astronomers to study the photometric structure of galactic
nuclei with unprecedented detail. Most studies have focused on
early-type galaxies and bulges, and many systems have now been imaged
(e.g., Crane \etal 1993; Jaffe \etal 1994; Lauer \etal 1995; Forbes,
Franx \& Illingworth 1995; Carollo \etal 1997, hereafter C97). The
main result is that at the $\sim 0.1''$ resolution limit of HST,
virtually all galaxies have power-law surface brightness cusps, $I
\propto r^{-\gamma}$, with $\gamma>0$ and no observed transition to a
homogeneous core. In addition, the surface brightness profiles fall in
two categories: `cores', which have a break at a resolved
radius,\footnote{The term `core' is somewhat confusing because of its
multiple meanings. Throughout this paper, the terms `core profile' and
`core galaxy' will refer to (a galaxy with) a cuspy core, $I \propto
r^{-\gamma}$ with $\gamma \leq 0.3$, and not to (a galaxy with) a
homogeneous core, $I = $ constant.} and `power-laws', which have no
clear break (Faber \etal 1997; hereafter F97). The former have $\gamma
\leq 0.3$, while the latter have $\gamma > 0.3$. The nuclear
properties correlate with luminosity: in general, galaxies with $\MV <
-22$ have core profiles, galaxies with $\MV > -20.5$ have power-law
profiles, and both profile types occur in galaxies with $-22 < \MV <
-20.5$. It is the aim of the present paper to provide a quantitative
framework for understanding these observations through models with
massive black holes (BHs).

Black holes are believed to be responsible for the energetic processes
in active galaxies and quasars (e.g., Peterson 1997; Fabian 1997).
Many or most quiescent galaxies may have been active in the past, and
may thus also contain BHs (So{\l}tan 1982; Chokshi \& Turner 1992).
It has been well documented theoretically that the presence of a BH in
the center of a stellar system induces a power-law cusp in the mass
distribution, $\rho \propto r^{-p}$ at asymptotically small radii,
with a corresponding cusp $I \propto r^{-q}$ in the projected surface
density (at asymptotically small radii $q = \max[0,p-1]$, but this
need not be true at observationally accessible radii; e.g., Gebhardt
\etal 1996). With globular clusters in mind, Bahcall \& Wolf (1976)
showed that $p={7\over 4}$ for a system with a BH that has been
present longer than the two-body relaxation time (this holds for equal
mass stars; the results with a mass spectrum are more complicated,
cf.~Bahcall \& Wolf 1977). In galaxies, the two-body relaxation time
typically exceeds the Hubble time, even in the central regions (Rauch
\& Tremaine 1997). Young (1980) inferred $p={3\over 2}$ for the case
of adiabatic BH growth (i.e., growth slow compared to the dynamical
time) in an isothermal sphere (see also Cipollina 1995). Quinlan,
Hernquist \& Sigurdsson (1995) showed that non-isothermal initial
conditions may yield other values of $p$. Adiabatic BH growth in
rotating models was studied by Lee \& Goodman (1989) and Cipollina \&
Bertin (1994). Sigurdsson, Hernquist \& Quinlan (1995) presented
$N$-body calculations of BH growth, and discussed the relation between
adiabatic invariance and the BH growth time.  Scenarios in which BHs
appear non-adiabatically have also been addressed. Stiavelli (1998)
showed that galaxy formation by violent relaxation around a
pre-existing BH yields $p={3\over 2}$, as in the models of Young
(1980).  Accretion of an external BH can either weaken an existing
cusp (Quinlan \& Hernquist 1997) or create a new cusp (Nakano \&
Makino 1997), depending on the assumed initial structure of the
accreting galaxy.

Observed power-law cusps need not necessarily be associated with BHs,
for several reasons. First, one of the simplest stellar dynamical
equilibrium systems is the singular isothermal sphere, which has $\rho
\propto r^{-2}$ and no BH (e.g., Binney \& Tremaine 1987). Second, a
power-law density cusp can easily be created in a stellar system
through dissipational processes (Mihos \& Hernquist 1994). Third,
cusps can form naturally without BHs even in absence of dissipation;
e.g., dissipationless simulations of galaxy formation produce dark
halos with $\rho \propto r^{-1}$ near the center (e.g., Navarro, Frenk
\& White 1996). Hence, the fact that all galaxies observed with HST
have surface brightness cusps does not necessarily imply that all
galaxies have BHs (Kormendy \& Richstone 1995).

We focus here on the simple models of adiabatic BH growth in a
homogeneous isothermal core proposed by Young (1980). These models
have been used to interpret HST photometry for two well-studied
galaxies that also have kinematically determined BH masses, namely M87
and M32. Given that observed surface brightness cusps can be
interpreted in so many ways, it is quite remarkable how successful the
models have been for these two galaxies.  The photometrically
determined BH mass for M87 (Young \etal 1978; Lauer
\etal 1992a) agrees within the errors to that determined kinematically
(Harms \etal 1994; Macchetto \etal 1997), as is the case for M32 (Lauer
\etal 1992b; van der Marel \etal 1998). 

The success of Young's models for individual galaxies justifies a more
detailed investigation of their viability. We therefore study whether
and how these models fit in with our general understanding of galactic
nuclei, as characterized by four important observational findings: (i)
surface brightness profiles have cusps; (ii) cusp slopes correlate
with luminosity; (iii) nuclear properties of early-type galaxies obey
scaling relations similar to those of the fundamental plane (Lauer
1985; Kormendy 1985; F97); and (iv) kinematically determined BH masses
in nearby galaxies tend to scale roughly with galaxy luminosity (e.g.,
Kormendy \& Richstone 1995; Ho 1998). The key questions that these
findings pose are: (a) is it possible that all surface brightness
cusps observed with HST, those in core profiles as well as those in
and power-law profiles, are due to adiabatic BH growth in homogeneous
cores?; (b) can this scenario naturally explain the observed
correlation between nuclear cusp slope and luminosity?; (c) are the
properties of the required homogeneous `progenitor' cores plausible in
view of the fundamental plane?; and (d) is the BH mass distribution
thus implied by HST photometry consistent with that suggested by
kinematical observations of nearby galaxies? The main result of the
present paper is that all four questions can be answered
affirmatively.

The paper is organized as follows. Section~\ref{s:models} presents a
general description and discussion of the models.
Section~\ref{s:trends} combines the models with established scaling
relations for early-type galaxies to obtain predictions for the
nuclear photometric properties of these galaxies as function of galaxy
luminosity. Section~\ref{s:indifit} presents model fits to HST
photometry of 46 individual galaxies. Section~\ref{s:conc} summarizes
and discusses the conclusions of the paper. Some additional galaxies
that did not pass the selection criteria of the main sample are
discussed in Appendix~\ref{s:AppA}.

A Hubble constant $H_0 = 80 \kms \Mpc^{-1}$ is used throughout this
paper. This does not directly influence the data-model comparison, but
does set the length, mass and luminosity scales in physical
units. Specifically, distance, length and mass scale as $H_0^{-1}$,
luminosity scales as $H_0^{-2}$, and mass-to-light ratio scales as
$H_0$. All quantities in this paper pertain to the photometric
{\VV}-band.

\section{The models}
\label{s:models}

\subsection{Calculations}
\label{ss:modcalc}

We study adiabatic BH growth in spherical models. The adiabaticity
implies, by definition, that the orbital actions (the radial action
$J_r$ and the angular momentum $L$) are conserved (e.g., Binney \&
Tremaine 1987). The dependence of the phase-space distribution
function on the actions is therefore identical before and after BH
growth. However, the BH growth does change the gravitational
potential, so the relation between the actions and the phase-space
coordinates is different in the initial and the final models. An
explicit expression for the phase-space distribution function $f$ of
the final model, as function of the energy $E$ and the angular
momentum $L$, can be obtained with an iterative algorithm (Young
1980). Given a trial estimate for the mass density $\rho$, an
iteration step proceeds as follows: use Poisson's equation to
calculate the gravitational potential generated by $\rho$, and add the
contribution of the BH; calculate the actions in this potential on a
grid in $(E,L)$ space; calculate the distribution function $f(E,L)$ on
the grid, using the fact that $f$ is the same function of the actions
as in the (pre-specified) initial model; calculate a new estimate for
$\rho$ by integration of $f(E,L)$ over velocity space. Iteration
proceeds until the mass-density has converged. Projected model
properties can be evaluated after convergence through straightforward
numerical integration. This algorithm was implemented by, e.g.,
Quinlan \etal (1995). Gerry Quinlan kindly provided his software for
use in this paper, and all models discussed here were calculated with
this software.

\subsection{Initial conditions}
\label{ss:inicond}

The assumption of adiabatic BH growth by itself is not very
informative. In principle, any observed surface brightness profile is
consistent with this assumption, for any arbitrary value of the
assumed current BH mass.\footnote{To obtain the appropriate initial
state, one merely needs to adiabatically remove the assumed BH from
the observed current state. The BH mass cannot be chosen completely
arbitrarily if one demands that the initial state is dynamically
stable and physically plausible (e.g., has a monotonically decreasing
density with radius), but these constraints by themselves are not
likely to strongly constrain the BH mass.} So to make progress one
must assume some knowledge about the properties of the initial model.

The analysis here is restricted to models in which the initial system
is a (non-singular) isothermal sphere, as in Young (1980). The model
predictions at large radii will not be used, so the only relevant
assumption in the present context is that the initial system has a
homogeneous isothermal core. Young's models have been widely discussed
in the literature, yield definite predictions, and have not previously
been compared in detail to HST results for a large sample of
galaxies. They are therefore the natural starting point for any
investigation. In addition, stellar systems do in fact exist for which
the central regions are well approximated by an isothermal
sphere. Globular clusters that have not undergone core collapse due to
two-body relaxation are observed to have constant density isothermal
cores (e.g., Trager, King \& Djorgovski 1995). This may indicate that
isothermal cores are a natural attribute of any stellar system. The
fact that constant density cores are not found in the centers of
early-type galaxies may indicate merely that the density distribution
in their nuclear regions has been altered by the presence of a BH, and
it is exactly this proposition that will be tested by the present
paper.

Young's models make two strong assumptions about the relation between
BHs and galaxy formation. It is assumed that galaxies form and relax
before BHs start to grow, and it is assumed that the subsequent BH
growth is slow compared to the dynamical time. These assumptions
provide a viable scenario (e.g., Murphy, Cohn \& Durisen 1991), but
are not unique. One alternative is that galaxy and BH formation occur
rapidly and simultaneously (e.g., Haehnelt \& Rees
1993). Interestingly, Stiavelli (1998) recently showed that galaxy
formation around a pre-existing BH, for a given BH mass, leads to a
similar end-state as adiabatic growth in a pre-existing homogeneous
core. This indicates that the predictions of Young's models may have a
larger range of applicability than strictly indicated by their
underlying model assumptions. In particular: (i) a good fit to the
data with Young's models does not necessarily imply that a BH indeed
grew adiabatically, or that the galaxy ever had a homogeneous core;
and (ii) growth of BHs in galaxies before or during galaxy formation
does not necessarily invalidate the predictions of Young's models.

\subsection{Predictions}
\label{ss:predic}

We define the dimensionless BH mass $\mu \equiv \Mbh / M_{\rm core}$,
where $M_{\rm core} \equiv {4\over 3}\pi \rho_0 r_0^3$ is a measure of
the mass of the initial isothermal core, and $r_0$ and $\rho_0$ are
the core radius and central density. The radius of the BH `sphere of
influence' is $\rbh \equiv G \Mbh / \sigma_0^2 = 3 \mu r_0$, where
$\sigma_0$ is the velocity dispersion of the initial isothermal sphere
($\sigma_0^2 = {4 \over 9} \pi G \rho_0 r_0^2$; Binney \& Tremaine
1987). The shape of the projected intensity profile after BH growth,
$I(r)$, depends only on $\mu$; the quantities $r_0$ and $\rho_0 r_0$
determine the radial and intensity scaling.  Figure~\ref{f:models}
shows $I(r)$ for various values of $\mu$.

\placefigure{f:models}

Before discussing the implications of the BH growth, it is worth
pointing out a peculiar property of the isothermal sphere that is
visible in Figure~\ref{f:models}. The logarithmic slope of the
intensity profile decreases monotonically from zero at asymptotically
small radii, to ${\rm d}\log I / {\rm d}\log r = -1.40$ at
$\log(r/r_0) = 0.34$ (indicated by a vertical dashed line in
Figure~\ref{f:models}). Here the intensity profile has an inflexion
point. The logarithmic slope then {\it increases} with radius and has
the value ${\rm d}\log I / {\rm d}\log r = -1$ at asymptotically large
radii. In real galaxies the intensity profile generally does not have
an inflexion point, and the logarithmic slope is significantly steeper
at large radii. In the remaining figures of this paper we replaced the
model predictions at $\log(r/r_0) > 0.34$ with a power-law of slope
${\rm d}\log I / {\rm d}\log r = -1.40$. This was done for visual
purposes only; the actual data-model comparison will always be
restricted to radii with $\log(r/r_0) < 0.34$. To model observed
profiles at radii with $\log(r/r_0) > 0.34$ would require initial
conditions that better reproduce the observed behavior of real
galaxies at large radii. This would certainly be useful, but we will
argue in Section~\ref{ss:fitresults} that this will not change the
main conclusions of this paper.

Figure~\ref{f:models} shows that the adiabatic BH growth yields the
well-known asymptotic slope $I \propto r^{-1/2}$ at small
radii. However, this asymptotic behavior does not apply until at least
$\log(r/r_0) \lta -2$. This region is generally inaccessible to
observations. Core radii do not often exceed $r_0 \approx 3''$
(cf.~Section~\ref{s:indifit} below), for which the $\sim 0.1''$
resolution limit of HST corresponds to $\log(r/r_0) = -1.48$. The
logarithmic intensity slope at this radius can either be smaller or
larger than the asymptotic slope, depending on the value of $\mu$.

To get a quantitative understanding of the observable properties of
the models, we have fitted the predicted intensity profiles in
Figure~\ref{f:models} with the so-called `nuker-law' parameterization:
\begin{equation}
  I(r) = \Ib \> 2^{{\beta-\gamma}\over\alpha} \> (r/\rb)^{-\gamma} \>
         [1 + (r/\rb)^\alpha]^{-{{\beta-\gamma}\over\alpha}} .
\label{nukerdef}
\end{equation}
This parameterization has been used successfully to describe the
nuclear intensity profiles of a large sample of galaxies (Byun \etal
1996, hereafter B96; F97). The intensity is $I(r) \propto r^{-\gamma}$
at small radii and $I(r) \propto r^{-\beta}$ at large radii, with the
transition occurring at the break radius $\rb$; the parameter $\alpha$
determines the sharpness of the break, and $\Ib$ is the intensity at
$\rb$. The model predictions were fit over the radial range $-1.48
\leq \log(r/r_0) \leq 0.34$ (which is bracketed by the dashed vertical lines 
in Figure~\ref{f:models}), which is a typical range that may be
observationally accessible. The fitted profiles are shown as dotted
curves. The fit parameters are shown as function of $\mu$ in
Figure~\ref{f:nukpar}.

\placefigure{f:nukpar}

The most interesting parameter is $\gamma$, which measures the slope
of the intensity profile at the smallest accessible radii. It
increases monotonically with $\mu$, from $\gamma=0$ for $\mu=0$ to
$\gamma = 1.10$ for $\mu = 3.29$ (the largest $\mu$ shown in
Figure~\ref{f:models}). This spans the entire range of values observed
with HST (F97). It may seem remarkable that the slope $\gamma$ for
large values of $\mu$ is {\it steeper} than the slope at
asymptotically small radii. The reason for this is that for large
values of $\mu$ the BH `sphere of influence' exceeds the initial core
size. The BH then grows in the region where initially $\rho
\propto r^{-2}$, which leads to a steeper intensity profile after BH
growth (Quinlan \etal 1995). In fact, when larger values of $\mu$ are
considered than those plotted in Figures~\ref{f:models}
and~\ref{f:nukpar}, the slope $\gamma$ eventually even reaches values
in excess of 2. Such models are probably less relevant for real
galaxies though, which have $\gamma \lta 1.1$.

The parameters $\rb$ and $\Ib$ of the nuker-law fit determine the
radial and intensity scale of the model. They are of the same order as
the scale parameters $r_0$ and $\rho_0 r_0$ of the initial
isothermal sphere, but differ more from these scale parameters as
$\mu$ grows larger. The parameters $\alpha$ and $\beta$ of the nuker
law determine the shape of the intensity profile at and outside the
break radius. The models have $\beta \approx 1.5$, independent of
$\mu$, while $\alpha$ varies between $\alpha=1.6$ and $\alpha=2.5$,
depending on $\mu$. These results are not inconsistent with typical
observed values (F97; see also Section~\ref{ss:fitresults} below).

\section{Predictions based on scaling relations}
\label{s:trends}

The results presented in Figure~\ref{f:nukpar} show that Young's
models of adiabatic BH growth can in principle generate profiles with
similar cusp slopes as observed in either core galaxies or power-law
galaxies. In this section we address the question whether this does in
fact occur for plausible BH masses, and whether there is a natural
explanation for the observed correlation between cusp slope and
luminosity. These questions are addressed in a general sense, using
established scaling relations for early-type galaxies. Detailed fits
to individual galaxies are presented in Section~\ref{s:indifit}.
  
\subsection{Scaling Relations}
\label{ss:scalinglaws}

Early-type galaxies lie on a plane in the space of their four primary
global characteristics, i.e., the effective radius $r_{\rm eff}$, the
average projected intensity inside the effective radius $I_{\rm eff}$,
the velocity dispersion $\sigma$ and the total luminosity $L$
(Djorgovski \& Davis 1987; Dressler \etal 1987). For core galaxies,
this so-called fundamental plane has an analog in terms of nuclear
properties (Lauer 1985; Kormendy 1985). The relevant quantities are
now the break radius $\rb$ and break intensity $\Ib$, instead of
$r_{\rm eff}$ and $I_{\rm eff}$. In the present context we are
interested in the dependence of $\rb$ and $\Ib$ on luminosity. For the
sample of F97:
\begin{equation}
  \log \rb = -10.02 + 1.15 \log L ; \qquad
  \log \Ib = 14.45 - 1.00 \log L  ,
\label{rbIbcorerel}
\end{equation}
where $\rb$ is in $\pc$, $\Ib$ is in the {\VV}-band and in units of
$\Lsun \pc^{-2}$, and $L$ is the total {\VV}-band luminosity in
$\Lsun$. The RMS scatter around these relations is $\sim 0.30$ dex and
$\sim 0.32$ dex, respectively. The relations~(\ref{rbIbcorerel}) are
projections of relations (with smaller scatter) in the
four-dimensional $(\rb,\Ib,\sigma,L)$ space.\footnote{The global
fundamental plane that connects the quantities $(r_{\rm eff},I_{\rm
eff},\sigma,L)$ is effectively a plane in a three-dimensional space,
since $L \equiv 2 \pi r_{\rm eff}^2 I_{\rm eff}$ by
definition. However, the relation that connects $(\rb,\Ib,\sigma,L)$
for core galaxies is a relation in a four-dimensional space, since
there is no a priori reason why $\rb$ and $\Ib$ should uniquely
determine $L$. The slope of the core fundamental plane is not exactly
identical to that of the global fundamental plane; e.g., $\Ib \rb^2
\propto L^{1.3}$, whereas $I_{\rm eff} r_{\rm eff}^2 \propto L$.}

Core galaxies have cusp slopes $\gamma \lta 0.3$. The hypothesis here
is that these cusps are due to adiabatic BH growth in isothermal
cores. Figure~\ref{f:nukpar} then implies that $\mu \leq 0.11$ for
core galaxies. The same figure shows that
\begin{equation}
   r_0        \approx 1.25 \> \rb , \qquad 
   \rho_0 r_0 \approx 0.79 \> \Upsilon \Ib ,
\label{obstoinicorerel}
\end{equation}
to within 5\% and 8\%, respectively, for all $\mu \leq 0.11$. The
quantity $\Upsilon$ is the {\VV}-band mass-to-light ratio of the
stellar population, which transforms the observed projected light
intensity ($I_b$) to a projected surface mass density ($\rho_0
r_0$). Combination of equations~(\ref{rbIbcorerel})
and~(\ref{obstoinicorerel}) yields for the `progenitors' of core
galaxies (i.e., before they grew BHs):
\begin{equation}
  \log r_0        = -9.92 + 1.15 \log L ; \qquad
  \log \rho_0 r_0 = 14.35 - 1.00 \log L + \log \Upsilon .
\label{rrhoisorel}
\end{equation}

Power-law galaxies have no well-defined breaks, and hence $\rb$ and
$\Ib$ reduce to fit parameters with little physical significance.  As
a result, there is no analog to the relations~(\ref{rbIbcorerel})
between $\rb$, $\Ib$ and $L$ for power-law galaxies. The hypothesis
that we will explore here is that both core galaxies and power-law
galaxies evolved from progenitors with isothermal cores, and that the
scale radius and density of the progenitor cores of both types of
systems obeyed equations~(\ref{rrhoisorel}). This scenario predicts
for core galaxies and power-law galaxies of the same luminosity that:
(i) differences in cusp slope must be due to differences in BH mass;
and (ii) there should be no differences in surface brightness at radii
$r \gg r_0$, where BHs do not have an impact. The latter prediction
can be verified directly. Figure~\ref{f:fiverzero} shows as function
of luminosity, for each of the galaxies in the sample of
Section~\ref{s:indifit}, the observed {\VV}-band surface brightness at
$r = 5 \overline{r_0}$; here $\overline{r_0}$ is defined to be the
progenitor core radius predicted by
equation~(\ref{rrhoisorel}). Indeed, core galaxies and power-law
galaxies of the same luminosity have the same surface brightness
outside the nuclear region. This is partly a restatement of the fact
that these galaxies also follow the same set of global fundamental
plane relations.

\placefigure{f:fiverzero}

A BH in the center of a galaxy induces a $v \propto r^{-1/2}$ cusp in
the rotation velocity or velocity dispersion of a tracer population
near the nucleus. High spatial resolution kinematical measurements can
therefore yield accurate BH mass determinations for individual
galaxies.  This area of research has seen much progress in recent
years, due to the identification of new suitable tracers (water masers
clouds; ionized gas disks), increased spatial solution (e.g., from
HST) and improved modeling techniques. Reviews of the various
techniques for kinematical BH detection, and of the current state of
knowledge can be found in Kormendy \& Richstone (1995), Ford
\etal (1998), Richstone (1998), van der Marel (1998) and Ho (1998).
The BH masses follow a roughly linear scaling relation between BH mass
and galaxy luminosity. For disk galaxies this relation involves the
bulge or spheroid luminosity, rather than the total luminosity, but
this can be ignored in the present study, which deals exclusively with
early-type galaxies. A recent collection of published BH mass
estimates yields (van der Marel 1998; see also
Figure~\ref{f:BHmassdist}b below):
\begin{equation}
  \log \Mbh = -1.96 + 1.00 \log L , 
\label{BHlumcor}
\end{equation}
with a RMS scatter of $\sim 0.59$ dex. We will use this relation here,
although it remains unclear to what extent this relation may be
influenced by selection bias. In particular, there may be a population
of galaxies with lower BH masses that have to date remained
undetected. Complete samples of kinematically well-studied galaxies
are required to test this.

The mass-to-light ratios of early type galaxies have a shallow
correlation with luminosity. Axisymmetric dynamical models for
spatially resolved kinematical profiles yield (van der Marel 1991;
Magorrian \etal 1998, hereafter M98):
\begin{equation}
  \log \Upsilon = -1.12 + 0.18 \log L , 
\label{Upscor}
\end{equation}
with a RMS scatter of $\sim 0.12$ dex.

Combination of equations~(\ref{rrhoisorel}), (\ref{BHlumcor})
and~(\ref{Upscor}) yields for the quantity $\mu \equiv \Mbh /
({4\over3}\pi \rho_0 r_0^3)$ of the adiabatic BH growth models:
\begin{equation}
  \log \mu = 4.03 - 0.48 \log L .
\label{muLcor}
\end{equation}
A crude estimate for the RMS scatter in this relation is obtained by
adding the scatter in the constituent quantities in quadrature (this
ignores correlations), which yields $0.80$ dex. The absolute magnitude
in the {\VV}-band, $\MV$, is related to the luminosity $L$ according
to $\MV = 4.83 - 2.5\log L$, so alternatively:
\begin{equation}
  \log \mu = 3.10 + 0.192 \MV .
\label{muMVcor}
\end{equation}

\subsection{Profile shapes at the distance of the Virgo cluster}
\label{ss:profshapes}

Equations~(\ref{rrhoisorel}) and~(\ref{muLcor}) predict $r_0$, $\rho_0
r_0$ and $\mu$ for any given luminosity $L$. This allows calculation
of the complete surface brightness profile predicted by the adiabatic
BH growth scenario, in physical units, for a given luminosity.

\placefigure{f:virgoprof}

As an example we consider the case of galaxies at the distance of the
Virgo cluster (the typical distance for most of the galaxies in the
F97 sample). Figure~\ref{f:virgoprof} shows the predicted profiles for
absolute magnitudes $\MV = -22.5, -21.5, \ldots, -17.5$. This range of
$\MV$ values corresponds roughly to that for the Virgo galaxies in the
F97 sample. The values of $\mu$ dictated by equation~(\ref{muMVcor})
range from $\mu=0.061$ for $\MV = -22.5$ to $\mu=0.55$ for $\MV =
-17.5$, which is always small enough that the BH growth leaves the
profile outside $r_0$ unaltered. At the distance of the Virgo cluster
$\log r_0{\rm [arcsec]} = -9.57 - 0.46\MV$, and the initial core
radius is below the $\sim 0.1''$ HST resolution limit for all $\MV >
-18.63$.\footnote{In fact, at the distance of the Virgo cluster a
galaxy with an isothermal core following relation~(\ref{rrhoisorel})
would only be classified as a core galaxy for $\MV > -19.07$;
cf.~Figure~\ref{f:virgonuk} below.} So for $\MV = -17.5$ and $\MV =
-18.5$, the profiles before and after BH growth (dashed and solid
curves, respectively) are identical and both reflect the adopted
initial model outside the core radius (in this case a power law of
logarithmic slope $-1.40$, see Section~\ref{ss:predic}).  As the
luminosity is increased, the initial core radius $r_0$ (indicated by a
dotted vertical line) shifts into the observable range. The profiles
inside $r_0$ have a power-law cusp, due to the BH. For $\MV = -19.5$
and $\MV = -20.5$, the transition from the outer power-law slope to
the inner power-law cusp is shallow enough that these profiles would
still be classified in the F97 scheme as power-law profiles, not core
profiles. However, the dimensionless BH mass $\mu$, and therefore also
the power-law cusp slope $\gamma$ (cf.~Figure~\ref{f:nukpar}),
decreases with luminosity, whereas $r_0$ increases with
luminosity. These effects combine to produce a pronounced break in the
profiles for $\MV = -21.5$ and $\MV = -22.5$, leading them to be
classified as core profiles. Figure~\ref{f:virgoprof} therefore shows
that the models naturally predict a transition from power-law profiles
to core profiles with increasing luminosity, as seen in HST
observations.

\subsection{Comparison to observed cusp properties}
\label{ss:compobs}

For a more quantitative assessment of the model predictions we have
fitted nuker-law parameterizations to the profiles predicted for
various absolute magnitudes. The most relevant parameters in the
present context are the central cusp slope $\gamma$ and the break
radius $\rb$. These are shown as function of absolute magnitude in
Figure~\ref{f:virgonuk}, using as before the distance of the Virgo
cluster.  Results are shown not only for the standard BH masses given
by equation~(\ref{BHlumcor}), but also for the cases of BH masses that
are three times smaller and three times larger than suggested by this
relation, as well as for the no-BH case (which corresponds to the
initial isothermal models).

\placefigure{f:virgonuk}

F97 define core galaxies as galaxies that have both $\gamma \leq 0.3$
{\it and} $\rb \geq 0.16''$. All other galaxies are classified as
power-law galaxies.\footnote{In atypical cases galaxies may have
$\gamma \leq 0.3$ and $\rb \geq 0.16''$, yet show no clear break.
This can occur when the best nuker-law fit invokes an anomalously low
$\alpha \lta 0.5$. Examples are NGC 4564 in the F97 sample, and NGC
4589 in the C97 sample. These galaxies are classified as power-law
galaxies.} F97 find that in general, galaxies with $\MV > -20.5$ have
power-law profiles, galaxies with $\MV < -22$ have core profiles, and
both profile types occur in galaxies with $-22 < \MV < -20.5$. The
model predictions presented here are remarkably consistent with these
observations. The heavy solid curves in Figure~\ref{f:virgonuk} show
that models with the standard BH masses of equation~(\ref{BHlumcor})
predict that galaxies at the distance of Virgo fainter than $\MV =
-21.2$ are power-law galaxies, and that galaxies brighter than $\MV =
-21.2$ are core galaxies. Both the sense and the absolute magnitude of
the predicted transition are the same as observed! The BH masses of
galaxies are likely to have an intrinsic scatter, and this naturally
explains the observed existence of both power-law galaxies and core
galaxies in the region around the transition magnitude.

Both the absence of core galaxies at sufficiently faint luminosities
and the absence of power-law galaxies at sufficiently bright
luminosities is generically reproduced by the models. However, to
reproduce that the cut-offs occur at $\MV = -20.5$ and $\MV = -22$
requires a certain amount of fine-tuning. It is reproduced if there
are no galaxies at $\MV > -20.5$ with BH masses smaller than $0.6$
times those given by equation~(\ref{BHlumcor}), and if there are no
galaxies at $\MV < -22$ with BH masses larger than $1.5$ times those
given by equation~(\ref{BHlumcor}). A situation in which this may
occur naturally is if the relation between $\log \Mbh$ and $\log L$
actually has a somewhat shallower slope than suggested by
equation~(\ref{BHlumcor}). The slope in equation~(\ref{BHlumcor})
itself is not ruled out, but the cutoffs at $\MV = -20.5$ and $\MV =
-22$ are then somewhat surprising given that BH masses inferred from
kinematical observations of nearby galaxies have a RMS scatter around
equation~(\ref{BHlumcor}) of a factor $\sim 4$. Such a scatter would
lead one to expect to find some power-law galaxies at $\MV < -22$ and
some core galaxies at $\MV > -20.5$. Instead, the F97 sample contains
none of the former; it contains two of the latter (M31 and NGC 4486B),
but these are both rather atypical (among other things, both have a
double nucleus; Lauer \etal 1993, 1996). Possibly, larger samples will
shed more light on the rigidity of the observed cutoffs at $\MV =
-20.5$ and $\MV = -22$.

Observations indicate not only that the cusp slopes of early-type
galaxies correlate with luminosity, but also that the distribution of
cusp slopes is bimodal (Gebhardt \etal 1996; F97; but see Merritt 1997
for an alternative view). The primary evidence for this is a lack of
galaxies with $\gamma = 0.4 \pm 0.1$. Observed core galaxies have
$\gamma \lta 0.3$ and observed power-law galaxies have $\gamma \approx
0.8 \pm 0.3$. This cannot be explained with the present scenario if
all galaxies have the same $\Mbh/L$ given by
equation~(\ref{BHlumcor}), which predicts that galaxies with $-20.0 <
\MV < -21.2$ produce cusp slopes $\gamma = 0.4
\pm 0.1$ (cf.~Figure~\ref{f:virgonuk}). Hence, 
a bimodal distribution in $\Mbh/L$ is required to reproduce the
observed bimodality in $\gamma$. For example, assume that there are
two populations of galaxies in the luminosity range $-22 < \MV <
-20.5$ where core and power-law galaxies overlap: galaxies that have
three times smaller BH masses than suggested by
equation~(\ref{BHlumcor}), and galaxies that have three times larger
BH masses than suggested by this equation. Figure~\ref{f:virgonuk}
shows that the former galaxies will all be core galaxies with $\gamma
\leq 0.2$ and that the latter will all be power-law galaxies with
$\gamma > 0.5$, reproducing the observed bimodality.

Evidence has been mounting in recent years that there are two types of
elliptical galaxies: luminous ellipticals with boxy isophotes that are
pressure supported by an anisotropic velocity distribution, and small
ellipticals with disky isophotes that are flattened by
rotation. Kormendy \& Bender (1996) suggested a formal subdivision of
the Hubble sequence based on these criteria. F97 found an almost
perfect correspondence between this bimodality in the global
properties of elliptical galaxies and the bimodality in the nuclear
cusp slopes. Galaxies with disky isophotes have power-law profiles and
galaxies with boxy isophotes have core profiles. The scenario
presented here explains this as a result of higher values of $\Mbh/L$
in disky galaxies than in boxy galaxies. The number of galaxies with
kinematically determined BH masses is too small at present to either
confirm or rule out this hypothesis directly. However, the observed
bimodality in both the global and nuclear properties of elliptical
galaxies implies that a bimodal distribution of $\Mbh/L$ may not be
unnatural. The formation of disky galaxies is believed to be
characterized by dissipation, either during galaxy formation or at the
time of the last major merger. This suggests the presence of large
amounts of gas, a significant fraction of which may have been used to
form or feed a BH. This provides a plausible reason why $\Mbh/L$ would
be larger in disky galaxies than in boxy galaxies.

\section{Fits to individual galaxies}
\label{s:indifit}

The results of Section~\ref{s:trends} show that the scenario of
adiabatic BH growth combined with established scaling relations
reproduces the generic nuclear photometric properties of early-type
galaxies.  However, in any sample of real galaxies one should expect
an intrinsic scatter in the relevant model parameters, $r_0$,
$\rho_0$, $\Mbh$, as well as in distance. So to further address the
plausibility of the proposed scenario we proceed by analyzing the
surface brightness profiles of individual galaxies.

\subsection{Sample selection}
\label{ss:sample}

The heart of the sample studied here is the F97 sample, which is
itself a collection of galaxies from various sources, including Lauer
\etal (1995), Jaffe \etal (1994) and various WFPC/GTO programs. To the
F97 sample we added those galaxies in B96 and C97 that are not already
contained in the F97 sample.  From the combined sample we removed the
few galaxies that are not classified as an early type (E, E/S0 or S0)
in the RC3 (de Vaucouleurs \etal 1991). We removed two more galaxies
for which only the profile classification is available (core or
power-law), without further information on the surface brightness
profile. The resulting sample has 71 galaxies, and contains the
majority of the nearby early-type galaxies that have been imaged to
date with HST.

BHs with plausible masses do not influence the surface brightness
profile at radii $r \gta r_0$, cf.~Sections~\ref{s:models}
and~\ref{s:trends}. So to obtain meaningful information for a given
galaxy, there must be at least a small radial range $r \lta r_0$ for
which the surface brightness profile is not degraded by the $\sim
0.1''$ resolution of HST. We therefore removed all galaxies from the
sample with $\overline{r_0} < 0.3''$, where as before $\overline{r_0}$
is for each galaxy defined to be the progenitor core radius predicted
by equation~(\ref{rrhoisorel}). This removes all low-luminosity
galaxies with $\MV > -19.7$. For these galaxies one generally expects
to resolve neither a clear signature of a core, nor of a BH
(cf.~Figure~\ref{f:virgoprof}).\footnote{Kinematical detection of BHs
in such low-luminosity galaxies is difficult as well. The BH sphere of
influence is $\rbh \equiv G \Mbh / \sigma_0^2 = 3 \mu r_0$
(cf.~Section~\ref{ss:predic}). Equations~(\ref{rrhoisorel})
and~(\ref{muLcor}) yield $\log \rbh = -5.41 + 0.67 \log L$, so at the
distance of the Virgo cluster $\rbh < 0.20''$ for all $\MV > -19.7$.}
The remaining sample contains 46 galaxies. Basic properties are listed
in Tables~\ref{t:sampleC} and~\ref{t:samplePL}, for core and power-law
galaxies, respectively.

\placetable{t:sampleC}
\placetable{t:samplePL}

\subsection{Fitting strategy}
\label{ss:fitting}

The surface brightness profiles of early-type galaxies are well fit by
the nuker-law parameterization given in equation~(\ref{nukerdef}),
over the radial range $0.1'' \leq r \leq 10''$. The typical RMS error
of a fit is $0.02$ mag (B96). Heavy solid curves in
Figures~\ref{f:indigalsC} and~\ref{f:indigalsPL} show the nuker laws
that best fit the data for all core and power-law galaxies in the
sample, respectively. The literature sources of the nuker-law
parameters are listed in Tables~\ref{t:sampleC}
and~\ref{t:samplePL}. C97 present both {\VV}-band and {\II}-band data,
but we used only the {\VV}-band data from their study for consistency
with F97 and B96 (which present only {\VV}-band data). For galaxies
contained in the samples of both F97 and C97 we used only the F97 data
(but the C97 data yield similar results, see below). The brightness
profiles in the figures are shown as function of the mean radius $r =
\sqrt{ab}$, where $a$ and $b$ are the semi-major and semi-minor axes
of an elliptical isophote. This choice is most appropriate for
comparison to the predictions of our spherical models.

\placefigure{f:indigalsC}
\placefigure{f:indigalsPL}

We chose to fit the adiabatic BH growth models to the best-fitting
nuker-law parameterizations, rather than to the actual surface
brightness data points.\footnote{For convenience, we will nonetheless
refer to `the model fit to the data'.}  This not only has the
advantage of simplicity, but also allows galaxies to be included in
the study for which the best-fitting nuker-law parameters have been
published, but not the actual surface brightness data points. The
accuracy to be gained by fitting the actual data points is only very
modest, given how well the data are described by a nuker law. This was
verified explicitly for the case of M87, for which the BH mass
inferred by fitting to the data directly differs by only $0.04$ dex
from the BH mass inferred by fitting to the nuker law parameterization.

The non-hatched regions in the panels of Figures~\ref{f:indigalsC}
and~\ref{f:indigalsPL} indicate for each galaxy the region over which
the models were fit to the data. This region always excludes radii $r
< 0.1''$, which are degraded by the HST point-spread function
(PSF). It also excludes radii with $\log (r/\overline{r_0}) > 0.3$,
for which the surface brightness profile is not expected to be
influenced by the possible presence of a BH
(cf.~Sections~\ref{s:models} and~\ref{s:trends}). So at these radii
one would merely be comparing the initial model to the data, which is
meaningless, because no effort was made to find an initial model that
accurately represents the large radii behavior of real galaxies. The
fit region was never allowed to extend beyond $10''$; the HST data do
not typically extend this far, and the nuker-law description of the
data breaks down at these radii.

The adiabatic BH growth models have three free parameters: $r_0$,
which sets the radial scale; $\rho_0 r_0 / \Upsilon$, which sets the
intensity scale; and $\mu$, which determines the BH mass. The
brightness profiles for core galaxies have three well-determined
characteristic quantities: $r_b$, the radius at which a break occurs;
$I_b$, the intensity at that radius; and $\gamma$, the inner cusp
slope. For core galaxies it is therefore possible to infer a unique
best-fitting value for each model parameter. By contrast, the
brightness profiles of power-law galaxies have no clear break and no
characteristic scale radius. This yields a degeneracy in the modeling,
in the sense that models with different $r_0$ and $\Mbh$ can provide
very similar fits to the data. To break this degeneracy, we assume for
power-law galaxies that $r_0$ is known a priori, and we fix it to the
value $\overline{r_0}$ predicted by equation~(\ref{rrhoisorel}). So
for power-law galaxies we merely determine the $\Mbh$ values implied
by the hypothesis that power-law galaxies had progenitor cores that
obeyed the same scaling relations as core galaxies, while for core
galaxies we actually determine both $r_0$ and $\Mbh$ from the data
without further assumptions.

A fit to an observed brightness profile yields only the ratio $\Mbh /
\Upsilon$, not the BH mass $\Mbh$ itself. To obtain the latter requires 
knowledge of the stellar mass-to-light ratio $\Upsilon$, which can
only be determined from kinematical data. Where available, $\Upsilon$
was used from M98. For those galaxies not in their sample, $\Upsilon$
was estimated from equation~(\ref{Upscor}).

\subsection{Accuracy of the inferred BH masses}
\label{ss:fitacc}

Five galaxies in the F97 sample are also contained in the C97
sample. This allows a direct estimate to be obtained of the typical
formal errors in the model parameters (which, by definition, are the
RMS differences among model parameters determined from independent
data sets). This is important, because our approach of fitting to
nuker laws rather than to data points precludes use of the normal
statistical procedures. For the five galaxies in common to the F97 and
C97 samples, the RMS difference in the inferred $\log \Mbh$ values is
$0.09$ dex. This includes the combined random errors from both photon
noise and image deconvolution, because F97 used Lucy deconvolved
images obtained with the First Wide Field and Planetary Camera
(WFPC1), while C97 used images obtained with the Second Wide Field and
Planetary Camera (which has internal optics to correct for the
spherical aberration of the HST primary mirror). 

Some of the sample galaxies have minor or patchy dust obscuration.
This should have little effect on the inferred BH masses, because the
literature sources from which we took the nuker law parameterizations
always excluded or corrected for dust regions where necessary. There
are no galaxies in the sample with extreme dust absorption across the
nucleus, such as NGC 4261 (Ferrarese, Ford \& Jaffe 1996) or NGC 7052
(van der Marel \& van den Bosch 1998), because F97 rejected such
galaxies from their sample.

The effects of additional photometric components in the sample
galaxies is more difficult to assess. Out of 16 power-law galaxies in
the sample, 4 galaxies have a nuclear stellar cluster and 9 galaxies
have a nuclear stellar disk. Such components are not common in core
galaxies. Nuclear stellar clusters were always excluded from the fit
by the literature sources from which we took the nuker law
parameterizations, and they should therefore not bias the BH masses
inferred here. However, this remains somewhat speculative as long as
the relation between the presence of a nuclear stellar cluster and a
possible BH is unknown. By contrast, none of the literature sources
from which we took the nuker law parameterizations corrected for or
subtracted nuclear stellar disks. It is not clear whether this would
in fact be desirable, but it is certainly useful to know how sensitive
the inferred BH mass is to the presence of a stellar disk. The galaxy
NGC 4570 has nuclear stellar disk (van den Bosch, Jaffe \& van der
Marel 1998; van den Bosch \& Emsellem 1998; Scorza \& van den Bosch
1998), and provides a useful test case in this context. F97 presented
the overall best-fitting nuker law for this galaxy, while Scorza \&
van den Bosch (1998) performed a disk-bulge decomposition and
presented only the best-fitting nuker law to the bulge component. When
modeled with the scenario presented here, the BH masses implied by the
nuker law fits of F97 and Scorza \& van den Bosch differ by $0.19$
dex. This is consistent with the arguments presented in both papers
that nuclear stellar disks are not the cause of the steep cusps in
power-law galaxies, and suggests that nuclear stellar disks are not a
major source of uncertainty in the BH masses inferred here for
power-law galaxies.

So the overall uncertainty in the inferred BH masses due to photon
noise, PSF deconvolution, dust, nuclear stellar clusters and nuclear
stellar disks is $0.2$--$0.3$ dex. Tables~\ref{t:sampleC}
and~\ref{t:samplePL} show that the models invoke a BH to fit the data
for all galaxies, and this small error estimate suggests that these
detections are highly significant (given the model assumptions).  One
may nonetheless wonder whether there are some galaxies in our sample
for which the models may also be consistent with the data for $\Mbh =
0$. Gebhardt \etal (1996) studied non-parametric deprojections of the
surface brightness profiles of a large sample of early-type galaxies
observed with HST.  A total of 24 galaxies from our sample of 46 are
included in their sample. In only two of these galaxies (NGC 1600 and
NGC 4889) can the existence of an isothermal core not be excluded at
the $0.1''$ resolution limit of HST. If these statistics are
representative for our sample as a whole, then we may expect that the
models could be consistent with $\Mbh = 0$ for $\sim 4$ galaxies in
our sample of 46. This is small enough not to influence the
conclusions in the following sections on the overall properties of the
sample.

\subsection{Fit results}
\label{ss:fitresults}

Dashed curves in Figures~\ref{f:indigalsC} and~\ref{f:indigalsPL} show
the best model fits to the data for all galaxies. Dotted curves show
the isothermal profiles before BH growth. The parameters of the
best-fitting models are listed in Tables~\ref{t:sampleC}
and~\ref{t:samplePL}. 

Figure~\ref{f:indigalsC} shows the core galaxies in the sample. The
brightness profiles for these galaxies are generally well fit by the
models; the average RMS residual is only $0.03$ mag/arcsec$^2$. In
some galaxies, the models fit the data even outside the region where
the fit was performed (the hatched region); examples are A1020, NGC
720, and IC 1459. However, this is not generally true, as witnessed
by, e.g., A2052, NGC 524 and NGC 4168. This is because outside the
break radius $\rb$, real galaxies show a larger variety of properties
than the models. The nuker-law parameter $\beta$ measures the
intensity profile slope outside $\rb$, and $\alpha$ determines the
sharpness of the break. The core galaxies in the sample have rather
broad distributions of $\alpha$ and $\beta$. The 16\% and 84\%
percentiles for $\alpha$ are $0.90$ and $3.57$; for $\beta$ they are
$1.05$ and $1.64$. By contrast, the models for core galaxies always
have $\alpha = 2.3 \pm 0.1$ and $\beta = 1.55 \pm 0.1$
(cf.~Figure~\ref{f:nukpar} for $\mu \leq 0.11$, $\gamma \leq 0.3$).
This is because outside $\rb$ they always resemble the isothermal
sphere. With more general initial conditions it should be possible to
fit the observed profiles of core galaxies also outside the break
radius $\rb$. In particular, one could study initial conditions with
homogeneous cores but a larger variety of outer profile properties. We
have not explored this, because the resulting BH masses would probably
not be very different. The properties of the initial model at $r \gta
r_0$ have little influence on the predicted distribution after BH
growth at $r \lta r_0$, unless the size $\rbh$ of the BH sphere of
influence actually extends significantly into the region $r >
r_0$. This is not the case, since $\rbh = 3 \mu r_0$, and $\mu <
{1\over 3}$ for the large majority of the galaxies in the sample.
 
Figure~\ref{f:indigalsPL} shows the power-law galaxies in the sample.
The fits for these galaxies are somewhat poorer than for the core
galaxies, due in part to the fact that $r_0$ was fixed a priori (so
that one less free parameter is available to optimize the fit). The
average RMS residual is $0.08$ mag/arcsec$^2$. The mismatch is
typically in the sense that the predicted profiles have slightly more
curvature than the observed profiles. However, the fits are still
reasonable, given the simple nature of the models. There are a few
galaxies for which the fit is acceptable even outside the fit-region,
e.g., NGC 1439 and NGC 3377, but these are exceptions. More typically,
the predictions fall below the observations at large radii. As for
core galaxies, this can probably be improved with a more general
choice of initial conditions.

Overall, the fits to the HST photometry are quite acceptable,
especially for core galaxies and somewhat less so for power-law
galaxies. However, the main test of the proposed scenario is not
whether individual profiles can be well fit, but rather whether the
required BH masses are plausible. Based on the general discussion in
Section~\ref{s:trends} we already expect this to be the case, but this
can now be tested in detail.

\subsection{The BH mass distribution}
\label{ss:BHdist}

Figure~\ref{f:BHmassdist}a shows the BH masses for all 46 sample
galaxies as function of the {\VV}-band luminosity $L$. Pearson's
linear correlation coefficient $r = 0.68$ indicates a statistically
significant correlation between $\log\Mbh$ and $\log L$. The dashed
line is the best linear least-squares fit for the sample as a whole,
\begin{equation}
  \log \Mbh = 0.76 + 0.75 \log L .
\label{BHdisfit}
\end{equation}
The slope of this line is somewhat smaller than unity, but this is not
highly significant given the limited $\log L$ range of the sample (the
formal error on the inferred slope is $0.12$). The RMS scatter around
the fit is $0.31$ dex. The solid line is the best-fitting line of unit
slope
\begin{equation}
  \log \Mbh = -1.83 + 1.00 \log L ,
\label{BHdisunifit}
\end{equation}
for which the RMS scatter is $0.33$ dex. 

\placefigure{f:BHmassdist}

Figure~\ref{f:BHmassdist}b shows a collection of kinematical BH
detections in nearby galaxies (adapted from van der Marel 1998, with
NGC 7052 added from van der Marel \& van den Bosch 1998). The
long-dashed line is the relation given by
equation~(\ref{BHlumcor}). In Section~\ref{ss:compobs} we showed that
many of the photometric properties of early-type galaxies observed
with HST are consistent with the presence of BHs with masses following
equation~(\ref{BHlumcor}). A comparison of Figure~\ref{f:BHmassdist}a
and~\ref{f:BHmassdist}b now confirms this directly. The BH mass
distribution implied by the adiabatic BH growth models is consistent
with that determined kinematically. The difference between
equations~(\ref{BHlumcor}) and~(\ref{BHdisunifit}) is only $0.13$ dex,
significantly smaller than the RMS scatter in either relation.

Several other suggestions from Section~\ref{ss:compobs} are neither
strongly confirmed nor strongly disproved here. First, it was argued
that power-law galaxies should have higher BH masses than core
galaxies, at a given luminosity. We find here that the core galaxies
in the sample have $\langle \log(\Mbh/L) \rangle = -1.91$ with a RMS
scatter of $0.29$, whereas the power-law galaxies have $\langle
\log(\Mbh/L) \rangle = -1.68$ with a RMS scatter of $0.34$. So on
average power-law galaxies do have larger BH masses than core
galaxies, but the difference is not highly significant (the errors on
the mean $\langle \log(\Mbh/L) \rangle$ are $0.05$ dex and $0.09$ dex,
respectively). Second, it was argued from the observed absence of core
galaxies at $\MV > -20.5$ and the observed absence of power-law
galaxies at $\MV < -22$ that the relation between $\log \Mbh$ and
$\log L$ may have a slope less than unity. This is confirmed by
equation~(\ref{BHdisfit}), but again not at high significance. Third,
it was argued that the absence of galaxies with intermediate slope
brightness profiles may indicate a bimodality in the BH mass
distribution. Figure~\ref{f:BHmassdist}a shows that core galaxies and
power-law galaxies do tend to occupy somewhat different different
regions in the $(\Mbh,L)$ plane, but there does not seem to be a very
clear bimodality. In particular, there is a region around $\log L =
10.3$ ($\MV = -20.9$) and $\log \Mbh = 8.6$ in which the two types of
galaxies overlap.

\subsection{Kinematically determined BH masses for individual galaxies}
\label{ss:BHkin}

The accuracy of the photometrically determined BH masses,
$M_{\bullet,{\rm adi}}$, can be further assessed by studying
individual galaxies with kinematically determined BH masses. Four
galaxies in the sample have securely determined BH masses,
$M_{\bullet,{\rm sec}}$, from HST spectroscopy
(cf.~Figure~\ref{f:BHmassdist}b), namely: NGC 3115 (Kormendy \etal
1996), NGC 3377 (Kormendy \etal 1998; Richstone 1998), NGC 3379
(Gebhardt \etal 1998) and NGC 4486 ($=$M87; Harms \etal 1994;
Macchetto \etal 1997).  These BH mass determinations are typically
believed to be accurate to $| \log \Delta M_{\bullet,{\rm sec}} | \lta
0.3$. Figure~\ref{f:BHcompare}a compares the photometrically and
kinematically determined BH masses for these four galaxies. The
agreement is very good: on average $\langle \log(M_{\bullet,{\rm
adi}}/M_{\bullet,{\rm sec}}) \rangle = 0.19$ with a RMS scatter of
$0.25$ dex. The agreement is no poorer for power-law galaxies than for
core galaxies. NGC 3115 has both a nuclear stellar cluster and a
nuclear stellar disk, so the agreement for this galaxy strengthens the
arguments of Section~\ref{ss:fitacc} that these components do not bias
the BH mass estimate. Overall, the results in
Figure~\ref{f:BHcompare}a provide additional credibility to the
interpretation of observed surface brightness cusps in both core
galaxies and power-law galaxies as due to adiabatic BH growth in
initially isothermal cores. Unambiguous detection of BHs in individual
galaxies will always require high spatial resolution kinematical data,
but Figure~\ref{f:BHcompare}a shows that it is justified to use
photometrically determined BH masses such as those in
Tables~\ref{t:sampleC} and ~\ref{t:samplePL} as a useful guide when
such kinematical data are not available.

\placefigure{f:BHcompare}

The other galaxies in Figure~\ref{f:BHmassdist}b with kinematically
determined BH masses do not pass the criteria of our sample, despite
the fact that HST photometry is available for some of them. We briefly
discuss these galaxies in Appendix~\ref{s:AppA}, and present some
detailed model fits for M32. The results for M32 are not inconsistent
with those presented here for the main sample.

M98 analyzed ground-based stellar kinematical observations of a large
sample of early-type galaxies using stellar dynamical models with
phase-space distribution functions of the form $f(E,L_z)$. These
models can be viewed as the axisymmetric generalizations of spherical
isotropic models. M98 showed that most of the galaxies in their sample
must have BHs if these distribution functions are correct. However, it
is unknown whether real galaxies do in fact have isotropic velocity
distributions. The range of possible velocity anisotropies in
equilibrium systems is large (e.g., Statler 1987), and any anisotropy
may have a significant impact on the inferred mass distribution
(Binney \& Mamon 1982). So the accuracy of the inferred BH masses is
determined entirely by the unknown accuracy of the underlying model
assumptions, as is the case for our photometric models. Our sample has
21 galaxies in common with the sample of
M98. Figure~\ref{f:BHcompare}b compares the photometrically determined
BH masses for these galaxies to the values $M_{\bullet,{\rm iso}}$
implied by the $f(E,L_z)$ models. The BH masses agree reasonably well
for nearby galaxies with distances $D < 30 \Mpc$ (i.e., twice the
distance of the Virgo cluster): on average $\langle
\log(M_{\bullet,{\rm adi}}/M_{\bullet,{\rm iso}}) \rangle = -0.09$
with a RMS scatter of $0.41$ dex. However, the results for galaxies
with $D > 30 \Mpc$ disagree by an order of magnitude: $\langle
\log(M_{\bullet,{\rm adi}}/M_{\bullet,{\rm iso}}) \rangle = -0.93$
with a RMS scatter of $0.42$ dex. The isotropic kinematical models
require larger BH masses than the photometric models. The galaxies at
$D > 30 \Mpc$ are all core galaxies; if they have the massive BHs
suggested by M98, then our models predict much steeper brightness
cusps than observed. Van der Marel (1998) showed that this discrepancy
can be resolved by assuming that core galaxies have mild radial
velocity anisotropy, either throughout the galaxy or only outside the
core region (for the case of an isotropic core). Such radial
anisotropy would be consistent with several detailed studies of core
galaxies (van der Marel 1991; Bender, Saglia \& Gerhard 1994; Merritt
\& Oh 1997; Rix \etal 1998; Gerhard \etal 1998; Gebhardt \etal 1998), 
and could be due, e.g., to violent relaxation during dissipationless
galaxy formation (van Albada 1982). Radial anisotropy would cause
isotropic models to overestimate the BH mass (or invoke a BH when
there is none). The BH masses inferred from isotropic models would be
most in error for distant galaxies, for which the BH sphere of
influence would be smaller than the spatial resolution of ground-based
kinematical observations. A BH mass overestimate of a factor $10$ for
a distant galaxy does not require an implausibly large velocity
anisotropy, as was shown by van der Marel (1998) for the case of NGC
1600 ($D = 50 \Mpc$).

\subsection{Model parameters}
\label{ss:modpar}

In Section~\ref{ss:scalinglaws} we derived scaling relations for the
parameters of the adiabatic BH growth models. These can now be
verified with the results obtained for individual
galaxies. Figure~\ref{f:modpar} shows the inferred parameters, i.e.,
the progenitor core radius $r_0$, the progenitor core intensity scale
$\rho_0 r_0 / \Upsilon$, and the dimensionless BH mass $\mu$, for all
sample galaxies as function of luminosity. The solid lines are the
predicted scaling relations from equations~(\ref{rrhoisorel})
and~(\ref{muLcor}).

\placefigure{f:modpar}

The inferred values for $r_0$ and $\rho_0 r_0 /\Upsilon$ closely
follow the predicted relations (but remember that for power-law
galaxies $r_0$ was fixed a priori). The inferred values of $\mu$ are
higher for power-law galaxies than for core galaxies, in agreement
with the predictions of Section~\ref{s:trends}. However, the inferred
values show a somewhat steeper dependence on luminosity than predicted
by equation~(\ref{muLcor}). This is because the latter equation is
based on the assumed linear dependence between $\Mbh$ and $L$ from
equation~(\ref{BHlumcor}), while it was shown in
Section~\ref{ss:BHdist} that the inferred BH masses imply a somewhat
shallower slope. The long-dashed line in the right panel of
Figure~\ref{f:modpar} shows the prediction obtained by combining
equations~(\ref{rrhoisorel}) and~(\ref{Upscor}) with
equation~(\ref{BHdisfit}). This line does reproduce the trend in the
inferred $\mu$ values adequately. Overall, the results for individual
galaxies validate the more general analysis of Section~\ref{s:trends}.

\section{Concluding remarks}
\label{s:conc}

\subsection{Summary}
\label{ss:summary}

We have studied a scenario for the nuclear structure of early-type
galaxies based on the assumption that galaxies have central BHs that
grew adiabatically in homogeneous isothermal cores, as first proposed
by Young (1980), and we have compared the predictions of this scenario
to the nuclear properties of early-type galaxies observed with
HST. The models reproduce and explain many of the observed nuclear
photometric properties of early-type galaxies remarkably well.  The
main results are the following.
\begin{enumerate} 
\item The models can fit the full range of cusp slopes found
at the smallest radii observable with HST, $I \propto r^{-\gamma}$
with $0 \leq \gamma \lta 1.1$, despite the fact that the models always
predict $I \propto r^{-1/2}$ at asymptotically small radii. The cusp
slope at observable radii is determined by the dimensionless BH mass,
$\mu = \Mbh / M_{\rm core}$, where $M_{\rm core} \equiv {4\over 3}\pi
\rho_0 r_0^3$ is a measure of the mass of the `progenitor' core.
Models with larger $\mu$ yield larger cusp slopes $\gamma$.
\item `Core galaxies' have surface brightness profiles with a clear break,
inside which $\gamma \leq 0.3$. The break radius $\rb$ and break
surface brightness $\Ib$ obey scaling relations similar to those of
the fundamental plane. The cusp slopes in these galaxies can be
explained as the result of BHs with small values of $\mu$: $\mu \lta
0.1$. The quantities $r_b$ and $I_b$ in models with small $\mu$
reflect the core radius $r_0$ and scale intensity $\rho_0 r_0$ of the
progenitor core (with proportionality constants of order unity). So
the observed scaling relations for $\rb$ and $\Ib$ are due to scaling
relations that governed the properties of the homogeneous progenitor
cores.
\item `Power-law galaxies' are galaxies with $\gamma > 0.3$. Their
surface brightness profiles have no clear break, and there are no
characteristic radius and intensity for these galaxies that obey
scaling relations similar to those for core galaxies. The steep cusps
and absence of a break in these galaxies can be explained as the
result of BHs with $\mu \gta 0.1$. It is attractive to assume that the
progenitor cores of power-law galaxies obeyed the same scaling
relations as the progenitor cores of core galaxies. This is plausible
in view of the fact that power-law galaxies and core galaxies also
follow the same fundamental plane relations, and it is consistent with
observed brightness profiles outside the central region.
\item At the small radii of interest, the surface brightness profiles 
of individual galaxies are well fit by the models. To fit the data at
all radii requires models with more realistic initial conditions at
large radii. The inferred BH mass distribution for a sample of 46
galaxies with available HST photometry suggests a roughly linear
correlation between BH mass and {\VV}-band galaxy luminosity, of the
form $\log \Mbh \approx -1.83 + \log L$ (RMS scatter $0.33$ dex).
This agrees with the average relation for the 18 nearby galaxies with
kinematically well-determined BH masses, to within $0.13$ dex (much
better than RMS scatter in either relation). For the four galaxies in
common to both samples, the individually determined BH masses agree to
within $\sim 0.25$ dex RMS.
\item In general, galaxies with $\MV < -22$ are observed to have core 
profiles and galaxies with $\MV > -20.5$ are observed to have
power-law profiles. Both profile types are found in galaxies with $-22
< \MV < -20.5$. In the models, two effects influence the predicted
brightness profile as function of luminosity. First, the core mass
$M_{\rm core}$ increases more steeply with luminosity, $M_{\rm core}
\propto L^{1.5}$ (F97), than does the BH mass, $\Mbh \propto L$.
Hence, $\mu \propto L^{-0.5}$, so that lower luminosity galaxies are
predicted to have steeper cusp slopes.\footnote{M98 suggest that $\Mbh
\propto M = \Upsilon L \propto L^{1.18}$, cf.~equation~(\ref{Upscor}).
This implies $\mu \propto L^{-0.32}$, which does not significantly
alter the argument presented here.} Second, the core radius $r_0$
decreases with decreasing luminosity (F97), such that any break
present is more difficult to observe in low-luminosity galaxies. At
the distance of the Virgo cluster, the models predict core profiles
for $\MV < -21.2$ and power-law profiles for $\MV > -21.2$, for
galaxies that obey all scaling relations without intrinsic
scatter. This naturally reproduces both the sense and the absolute
magnitude of the observed transition. Intrinsic scatter in the
characteristic quantities $r_0$, $\rho_0 r_0$ and $\Mbh$ can explain
why both types of galaxies are observed in the region around the
transition magnitude.
\item Observations indicate a bimodality between core galaxies 
and power-law galaxies. The former have $\gamma \leq 0.3$ and the
latter have $\gamma = 0.8 \pm 0.3$, with few galaxies showing
intermediate slopes, $\gamma = 0.4 \pm 0.1$. This bimodality
correlates with the bimodality in the global properties of elliptical
galaxies. Core galaxies are luminous, have boxy isophotes and are
pressure supported by an anisotropic velocity distribution; power-law
galaxies are less luminous, have disky isophotes and are flattened by
rotation. The scenario presented here explains the observed bimodality
in cusp slopes as a result of higher values of $\Mbh/L$ in disky
galaxies than in boxy galaxies. The presence of gas and dissipation
during the formation or last major merger of disky galaxies may
provide a natural explanation for this.
\end{enumerate}

\subsection{Discussion}
\label{ss:disc}

The scenario presented here is not unique. Observed surface brightness
cusps can be interpreted in many different ways, as stressed in
Section~\ref{s:intro}. For example, the observed cusps may have formed
dissipationally during galaxy formation, and may not have any relation
to BHs at all. Our scenario is also oversimplified in that it ignores
the important issue of accretion events and mergers between galaxies,
which will almost certainly alter the brightness profiles of the
constituent galaxies (F97). Nonetheless, the proposed scenario
naturally predicts observational results that would otherwise remain
unexplained (e.g., the correlation between cusp slope and luminosity),
and it yields BH masses that agree with those determined
kinematically. These findings single out the present scenario as a
very attractive one among many possible ones, but it remains
surprising that such a simple scenario would work so well.

The success of Young's scenario does not teach us much about the
processes of BH and galaxy formation, and their connection. Even if
the surface brightness profiles predicted by the models for a given BH
mass are correct, this still does not imply that the BHs in galaxies
must have grown adiabatically, or that there ever were progenitor
systems with homogeneous cores. Stiavelli (1998) showed that violent
relaxation of stars around a pre-existing BH yields a similar
end-state as do the models of Young (1980). A wide variety of
scenarios in which BHs form before, during, or after the galaxies in
which they reside, may therefore all lead to a similar end-state. In
fact, this may well be the key to explaining why Young's models fit
the observations so well, despite their apparent simplicity.

If galaxies did in fact start out with homogeneous cores, this still
leaves open the questions why these homogeneous cores formed in the
first place, and why they followed the scaling relations of
equation~(\ref{rrhoisorel}). However, the scenario discussed here does
resolve one puzzling issue. It explains through the presence of BHs
why the observed nuclear properties of core galaxies follow relations
like those of the fundamental plane, whereas the observed nuclear
properties of power-law galaxies don't. This reduces the search for
understanding of the global and nuclear parameter relations of
galaxies to a single problem, and suggests that our understanding of
the global fundamental plane relations of early-type galaxies (e.g.,
Bender, Burstein \& Faber 1992; Burstein \etal 1997) applies to the
nuclear parameter relations as well. For example, the fact that the
inferred progenitor cores of low-luminosity galaxies are smaller and
denser than those of high-luminosity galaxies can probably be
attributed to dissipation during galaxy formation (F97 and references
therein).

Possibly most importantly, the results in this paper provide strong
new support to the hypothesis that all galaxies have BHs, and that BH
masses scale approximately with galaxy luminosity according to
equations~(\ref{BHlumcor}) or~(\ref{BHdisunifit}). This remains
difficult to prove in a model independent way, but is also indicated
by several other studies of active and quiescent galaxies (e.g.,
Kormendy \& Richstone 1995; M98; Ho 1998). It is also consistent with
quasar statistics (So{\l}tan 1982; Chokshi \& Turner 1992; F97). If
every galaxy spheroid harbors a BH that was formed in a quasar phase
through matter accretion with efficiency $\epsilon$, then
equation~(\ref{BHdisunifit}) implies that $\epsilon = 0.04$ (van der
Marel 1998).


\acknowledgments

I thank Gerry Quinlan for kindly allowing me to use his adiabatic BH
growth software, and Tim de Zeeuw and Marcella Carollo for a careful
reading of the manuscript. This work was supported by an STScI
Fellowship, awarded by the Space Telescope Science Institute which is
operated by the Association of Universities for Research in Astronomy,
Incorporated, under NASA contract NAS5-26555.

\clearpage


\appendix

\section{Galaxies with kinematically determined BH masses}
\label{s:AppA}

Only four of the galaxies with kinematically determined BH masses in
Figure~\ref{f:BHmassdist}b are part of the photometric sample
discussed in Section~\ref{s:indifit}, namely NGC 3115, NGC 3377, NGC
3379 and NGC 4486 ($=$M87). The other 14 galaxies do not pass the
selection criteria of the sample discussed in Section~\ref{ss:sample}.

Our own Galaxy, M31, NGC 1068, NGC 4258, NGC 4594 and NGC 4945 are not
early-type galaxies. Apart from this, no HST photometry has been
published for NGC 4594 (Sombrero), while dust obscuration is a
significant problem in NGC 1068, NGC 4258 and NGC 4945. A useful
surface brightness profile is available only for M31, but there the
presence of a double nucleus (Lauer \etal 1993) prevents a meaningful
comparison between our models and the observed surface brightness
profile. The galaxies M84, NGC 4261, NGC 7052 and NGC 6251 are
elliptical galaxies, but they all have pronounced dust disks across
the nucleus. The rotation velocities of the ionized gas in these disks
have yielded the BH mass determinations for these galaxies, but the
dust prevents a reliable determination of the stellar brightness
profile. For Arp 102B no HST imaging has been obtained to date.

This leaves M32, NGC 4342 and NGC 4486B, for which good HST surface
brightness profiles are available. These galaxies are not part of our
photometric sample because they have $\overline{r_0} < 0.3''$ (where
$\overline{r_0}$ is the progenitor core radius predicted by
eq.~[\ref{rrhoisorel}]). Specifically, $\overline{r_0} = 0.22''$ for
M32, $\overline{r_0} = 0.09''$ for NGC 4342 and $\overline{r_0} =
0.03''$ for NGC 4486B. Thus, if these galaxies follow the scaling
relations of Section~\ref{s:trends}, one expects to resolve neither a
clear signature of a core, nor of a BH. The fact that BHs have in fact
been detected kinematically implies that these galaxies apparently do
not follow the scaling relations obeyed by other galaxies. For M32 and
NGC 4342 we discuss this in some detail below. The case of NGC 4486B
was already discussed by F97. This satellite galaxy of M87 has a
brightness profile with a core, despite its low luminosity ($\MV =
-17.57$). F97 suggest that it was once more massive, but was stripped
of mass through interaction with M87. NGC 4486B also has a closely
separated double nucleus (Lauer \etal 1996), which prevents any
meaningful comparison between our models and the observed surface
brightness profile.

\subsection{M32}
\label{ss:AppM32}

M32 is especially interesting in the present context because Lauer
\etal (1992b) showed that HST photometry for M32 can be adequately
interpreted in terms of Young's adiabatic BH growth models. M32 is
classified by F97 as a power-law galaxy. In general we have used the
approach for power-law galaxies of fixing $r_0$ to the value predicted
by equation~(\ref{rrhoisorel}) when doing the adiabatic BH growth
model fit. However, the predicted $\overline{r_0} = 0.22''$ may not be
appropriate for M32. The scaling relation in
equation~(\ref{rrhoisorel}) is based on the core galaxies in our
sample, which span the luminosity range $10.15 \leq \log L \leq
11.35$. By contrast, M32 has $\log L = 8.57$. Small uncertainties in
the slope of equation ~(\ref{rrhoisorel}) therefore translate to large
uncertainties in the predicted $r_0$ for M32. Based on these
considerations we have done a model fit for M32 in which $r_0$ was not
fixed a priori, but was varied to optimize the fit. The fit itself was
done over the range of radii with $r \geq 0.1''$ and $\log (r/r_0) <
0.3$, where the upper limit of the fit range was adjusted iteratively
during the fit.

\placetable{t:sampleM}
\placefigure{f:indigalsM}

The best fit is shown in Figure~\ref{f:indigalsM}; its parameters are
listed in Table~\ref{t:sampleM}. The inferred BH has $\log \Mbh =
6.68$, which agrees with the kinematically determined value $\Mbh =
6.59$ (from van der Marel \etal 1998, modified to the same distance
used here) to within $0.09$ dex. This agreement is similar as for the
galaxies with kinematically well-determined BH masses shown in
Figure~\ref{f:BHcompare}. The value of $\Mbh/L$ for M32 is not
atypical, and falls exactly on the relation of
equation~(\ref{BHlumcor}). However, the best-fit model has $r_0 =
0.81''$, which exceeds the prediction of equation~(\ref{rrhoisorel})
by $0.57$ dex. This is approximately twice the RMS scatter around
equation~(\ref{rrhoisorel}) for core galaxies, insufficient to provide
strong evidence that the progenitor core of M32 violates the scaling
relations inferred for higher-luminosity galaxies. First, the
discrepancy is only at the $\sim 2\sigma$ level; second, uncertainties
in the slope of equation~(\ref{rrhoisorel}) may have exaggerated the
discrepancy; and third, the luminosity of M32 may have decreased over
time due to interactions with M31, with resulting stripping of
material. Overall, the results for M32 are not inconsistent with the
general scenario discussed in the context of our main sample.

\subsection{NGC 4342}
\label{ss:AppN4342}

NGC 4342 is an S0 galaxy in Virgo with a nuclear stellar disk (van den
Bosch \etal 1998; Scorza \& van den Bosch 1998). It has a power-law
brightness profile, as expected for its luminosity ($\MV=-18.49$). NGC
4342 violates the scaling relations of Section~\ref{s:trends} by
having an anomalously large $\Mbh/L$.  Cretton \& van den Bosch (1998)
use HST stellar kinematics to derive $\Mbh = 3.0 \times 10^8
\Msun$. This yields $\log \Mbh/L = -0.63$, which exceeds the average
relation in equation~(\ref{BHlumcor}) by $1.33$ dex. This contrasts
with the case of M32, which has a `normal' $\Mbh/L$, but a larger
$r_0$ than expected.

To model the NGC 4342 surface brightness profile one would like to
assume that it had a progenitor core obeying
equation~(\ref{rrhoisorel}), i.e., $r_0 = 0.09''$, but that it somehow
grew an anomalously large BH. Interpretation of the observed
brightness profile then requires an adequate model of BH growth that
influences the region {\it outside} the homogeneous progenitor
core. For this one needs an initial model that reproduces the large
radii behavior of real galaxies. As discussed in
Section~\ref{ss:predic}, the isothermal sphere is not sufficient for
this. Young's models therefore do not allow us to address whether of
not NGC 4342 is consistent with the adiabatic BH growth scenario.


\ifsubmode\else
\baselineskip=10pt
\fi


\clearpage


\ifsubmode\else
\baselineskip=14pt
\fi


\newcommand{\figcapmodels}{Heavy solid curves are the predicted 
intensity profiles $I(r)$ for models of adiabatic BH growth in an
isothermal sphere. The dimensionless BH mass for the models is, from
bottom to top, $\mu = 0$, $0.011$, $0.034$, $0.11$, $0.34$, $1.05$ and
$3.29$, respectively. The rightmost of the two dashed vertical lines
marks the radius at which the initial model has an inflexion point.
The other dashed vertical line marks the radius that corresponds to
the $0.1''$ resolution limit of HST for a case in which $r_0 = 3''$.
Dotted curves are nuker laws fitted to the models over the radial
range between the vertical lines. The quantities $r_0$ and $\rho_0$
are the core radius and central density of the initial
model.\label{f:models}}

\newcommand{\figcapnukpar}{Parameters of nuker-law fits to the predicted 
intensity profiles for models of adiabatic BH growth in an
isothermal sphere, as function of the dimensionless BH mass $\mu$.
Each fit was performed over the radial range indicated in
Figure~\ref{f:models}. The quantities $r_0$ and $\rho_0$ are the core
radius and central density of the initial model.\label{f:nukpar}}

\newcommand{\figcapfiverzero}{Observed {\VV}-band surface 
brightness $\Sigma_V$ at the radius $r = 5 \overline{r_0}$, as
function of {\VV}-band luminosity $L$, for each of the galaxies in the
sample discussed in Section~\ref{ss:sample}. The radius
$\overline{r_0}$ is the progenitor core radius before BH growth
predicted by equation~(\ref{rrhoisorel}). Core galaxies and power-law
galaxies are indicated by open and closed symbols, respectively. Both
types of galaxies follow one and the same relation, indicating that
their surface brightness profiles outside the nuclear region are
similar.\label{f:fiverzero}}

\newcommand{\figcapvirgoprof}{Predicted {\VV}-band surface brightness 
profiles $\Sigma_V(r)$ for galaxies of different absolute magnitude,
at the distance of the Virgo cluster. The parameters $r_0$ and $\rho_0
r_0$ for each model were obtained from the scaling relations given in
the text, which are based on observations of large samples of
galaxies. Dashed curves show the model profiles before BH
growth. Heavy solid curves show the model profiles after adiabatic
growth of a BH with mass given by equation~(\ref{BHlumcor}). Each
panel shows the radial range from $0.1''$ to $10''$, which is the
range typically accessible with HST. A dotted vertical line in each
panel indicates the core radius $r_0$ of the initial model. This
radius is too small to be observable ($r_0 < 0.1''$) for $\MV = -18.5$
and $\MV = -17.5$. The models naturally explain the observed
transition from core profiles to power-law profiles with decreasing
luminosity.\label{f:virgoprof}}

\newcommand{\figcapvirgonuk}{Nuker-law parameters for model galaxies of 
different absolute magnitude, at the distance of the Virgo
cluster. The models assume adiabatic BH growth and the scaling
relations given in the text. The top panel shows the central cusp
slope $\gamma$, the bottom panel the break radius $\rb$. The
parameterizations were fit over the radial range between $0.1''$ and
$10''$, which is the range typically accessible with HST. Heavy solid
curves show the predictions for models with BH masses given by
equation~(\ref{BHlumcor}) (corresponding to the heavy solid profiles
in Figure~\ref{f:virgoprof}). Dashed curves show the predictions for
the initial isothermal models, i.e., for models without BHs
(corresponding to dashed profiles in Figure~\ref{f:virgoprof}). The
long-dashed curves show the predictions for models that have three
times smaller BH masses than indicated by equation~(\ref{BHlumcor});
the dash-dot curves show the predictions for models that have three
times larger BH masses than indicated by this equation. A dotted
horizontal line in the top panel indicates the boundary between core
galaxies and power-law galaxies. A galaxy is defined to be a core
galaxy only if it has both $\gamma \leq 0.3$ {\it and} $\log \rb \geq
-0.8$ ($\rb \geq 0.16''$); it is a power-law galaxy otherwise
(F97).\label{f:virgonuk}}

\newcommand{\figcapindigalsC}{{\VV}-band surface brightness profiles for 
the core galaxies in the sample. The heavy solid curve for each galaxy
is the nuker law that best fits the HST photometry. Dashed curves are
the best-fitting adiabatic BH growth models. Dotted curves show the
isothermal brightness profiles of the models {\it before} BH
growth. Dashed curves are generally invisible at small radii, because
they overly the solid curves (indicating a good fit). Dotted curves
are generally invisible at large radii, because they overly the dashed
curves (indicating that the BH has no effect at large radii). To
obtain the dashed profiles from the dotted profiles through adiabatic
BH growth requires the BH masses listed in Table~\ref{t:sampleC}. The
brightness profiles in the hatched regions are uninfluenced by the BH
growth, and were excluded from the fit. In these regions the models
often represent the data poorly because an isothermal sphere is known
to be a poor approximation to the large radii behavior of real
galaxies. The fits at large radii can be improved by studying models
with more general initial conditions, but this is expected to have
little effect on the inferred BH masses.\label{f:indigalsC}}

\newcommand{\figcapindigalsPL}{As Figure~\ref{f:indigalsC}, but now for the 
power-law galaxies in the sample.\label{f:indigalsPL}}

\newcommand{\figcapBHmassdist}{(a) BH mass distribution as function of
{\VV}-band galaxy luminosity for the sample of early-type galaxies in
Tables~\ref{t:sampleC} and~\ref{t:samplePL}, inferred from adiabatic
BH growth models for HST photometry. Core galaxies and power-law
galaxies are indicated by open and closed symbols, respectively. The
dashed line is the best least-squares fit to the data. The solid line
is the best fit line of unit slope. Vertical dotted lines indicate the
luminosities corresponding to $\MV = -20.5$ and $\MV =-22$. (b) BH
mass distribution as function of {\VV}-band {\it spheroid} luminosity
for nearby galaxies with kinematically detected BHs (adapted from
figure~1a of van der Marel~1998, which also lists the references for
the individual galaxies). These $\Mbh$ values are typically believed
to be accurate to $|\Delta \log \Mbh | \lesssim 0.3$. The symbol type
indicates the kinematical tracer used to measure the BH mass:
($\odot$)~ionized gas kinematics of nuclear disks; ($\otimes$)~stellar
kinematical studies; ($\times$)~radio observations of water masers;
($\ast$)~time variability of broad double-peaked Balmer lines. The
long-dashed line is the best fit line of unit slope; it is consistent
with the solid line in the left panel to within $0.13$
dex.\label{f:BHmassdist}}

\newcommand{\figcapBHcompare}{Comparison of BH masses $M_{\bullet,{\rm adi}}$
inferred from the adiabatic BH growth models (shown along the
ordinate) to kinematically determined BH masses. The left panel shows
galaxies with kinematically determined BH masses $M_{\bullet,{\rm
sec}}$ that are `securely' known from HST spectroscopy. The right
panel shows galaxies for which BH masses $M_{\bullet,{\rm iso}}$ were
estimated from isotropic stellar dynamical models for ground-based
stellar kinematics (M98). All BH masses are in $\Msun$. The accuracy
of the $M_{\bullet,{\rm sec}}$ values is believed to be better than
$0.3$ dex. The formal errors in $M_{\bullet,{\rm adi}}$ and
$M_{\bullet,{\rm iso}}$ are small (typically $\lta 0.10$ dex), but the
accuracy of these estimates is determined entirely by the unknown
accuracy of the underlying model assumptions. Different symbols
indicate core galaxies closer than $30 \Mpc$, core galaxies further
than $30 \Mpc$, and power-law galaxies, respectively. The solid lines
indicate where the BH masses from different methods would agree
perfectly.\label{f:BHcompare}}

\newcommand{\figcapmodpar}{Adiabatic BH growth model parameters 
for the sample galaxies as function of {\VV}-band galaxy
luminosity. Shown from left to right are the progenitor core radius
$r_0$ in pc, the progenitor core intensity scale $\rho_0 r_0 /
\Upsilon$ in mag/arcsec$^2$, and the dimensionless BH mass $\mu$. The
solid lines are the predicted scaling relations from
equations~(\ref{rrhoisorel}) and ~(\ref{muLcor}). The core radius
$r_0$ for power-law galaxies was not obtained from a fit to the data,
but was fixed a priori to lie on the solid line, as described in the
text. The long-dashed line in the right panel shows the scaling
relation obtained by combining equations~(\ref{rrhoisorel})
and~(\ref{Upscor}) with equation~(\ref{BHdisfit}). In the scenario
discussed here, the homogeneous progenitor cores of core galaxies and
power-law galaxies follow a single set of nuclear scaling
relations.\label{f:modpar}}

\newcommand{\figcapindigalsM}{Observed and model surface brightness profiles
for M32, with line types as in
Figure~\ref{f:indigalsC}.\label{f:indigalsM}}


\ifsubmode
\figcaption{\figcapmodels}
\figcaption{\figcapnukpar}
\figcaption{\figcapfiverzero}
\figcaption{\figcapvirgoprof}
\figcaption{\figcapvirgonuk}
\figcaption{\figcapindigalsC}
\figcaption{\figcapindigalsPL}
\figcaption{\figcapBHmassdist}
\figcaption{\figcapBHcompare}
\figcaption{\figcapmodpar}
\figcaption{\figcapindigalsM}

\clearpage
\else\printfigtrue\fi

\ifprintfig


\clearpage
\begin{figure}
\centerline{\epsfbox{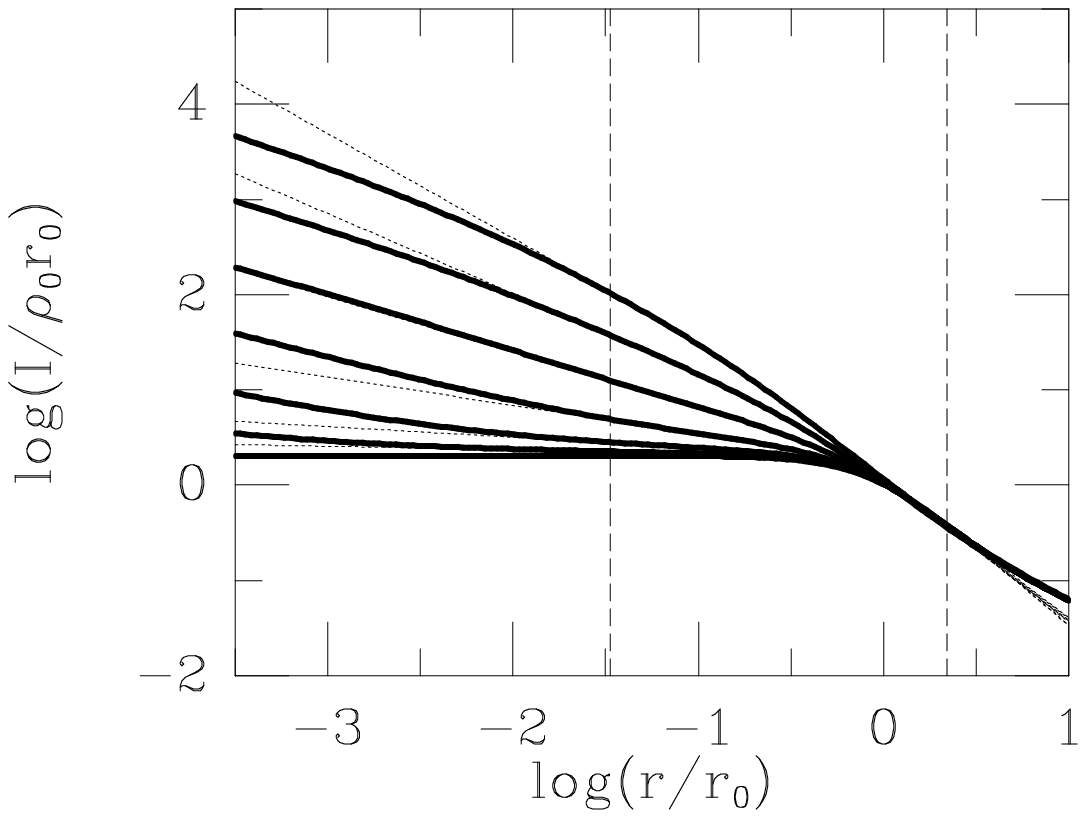}}
\ifsubmode
\vskip3.0truecm
\setcounter{figure}{0}
\addtocounter{figure}{1}
\centerline{Figure~\thefigure}
\else\figcaption{\figcapmodels}\fi
\end{figure}


\clearpage
\begin{figure}
\epsfxsize=16.0truecm
\centerline{\epsfbox{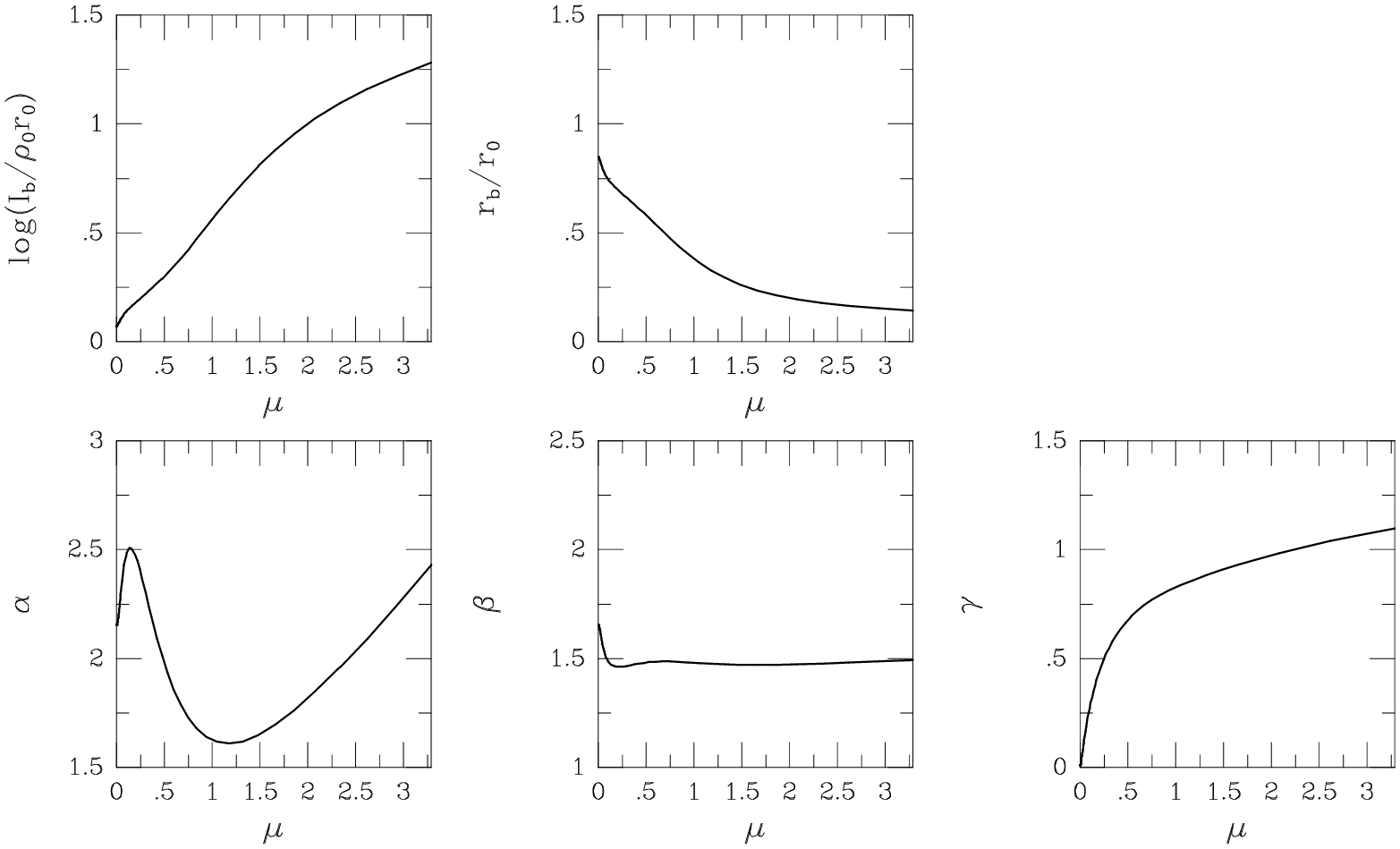}}
\ifsubmode
\vskip3.0truecm
\addtocounter{figure}{1}
\centerline{Figure~\thefigure}
\else\figcaption{\figcapnukpar}\fi
\end{figure}


\clearpage
\begin{figure}
\centerline{\epsfbox{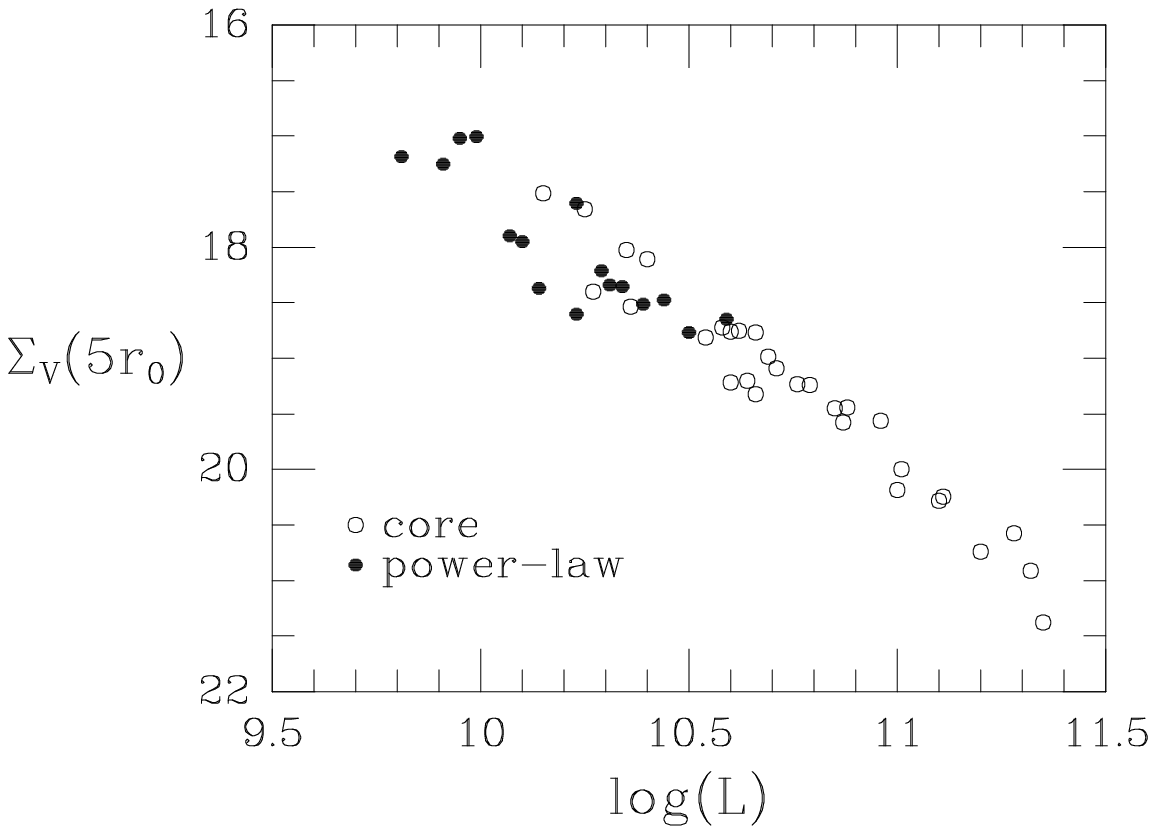}}
\ifsubmode
\vskip3.0truecm
\addtocounter{figure}{1}
\centerline{Figure~\thefigure}
\else\figcaption{\figcapfiverzero}\fi
\end{figure}


\clearpage
\begin{figure}
\centerline{\epsfbox{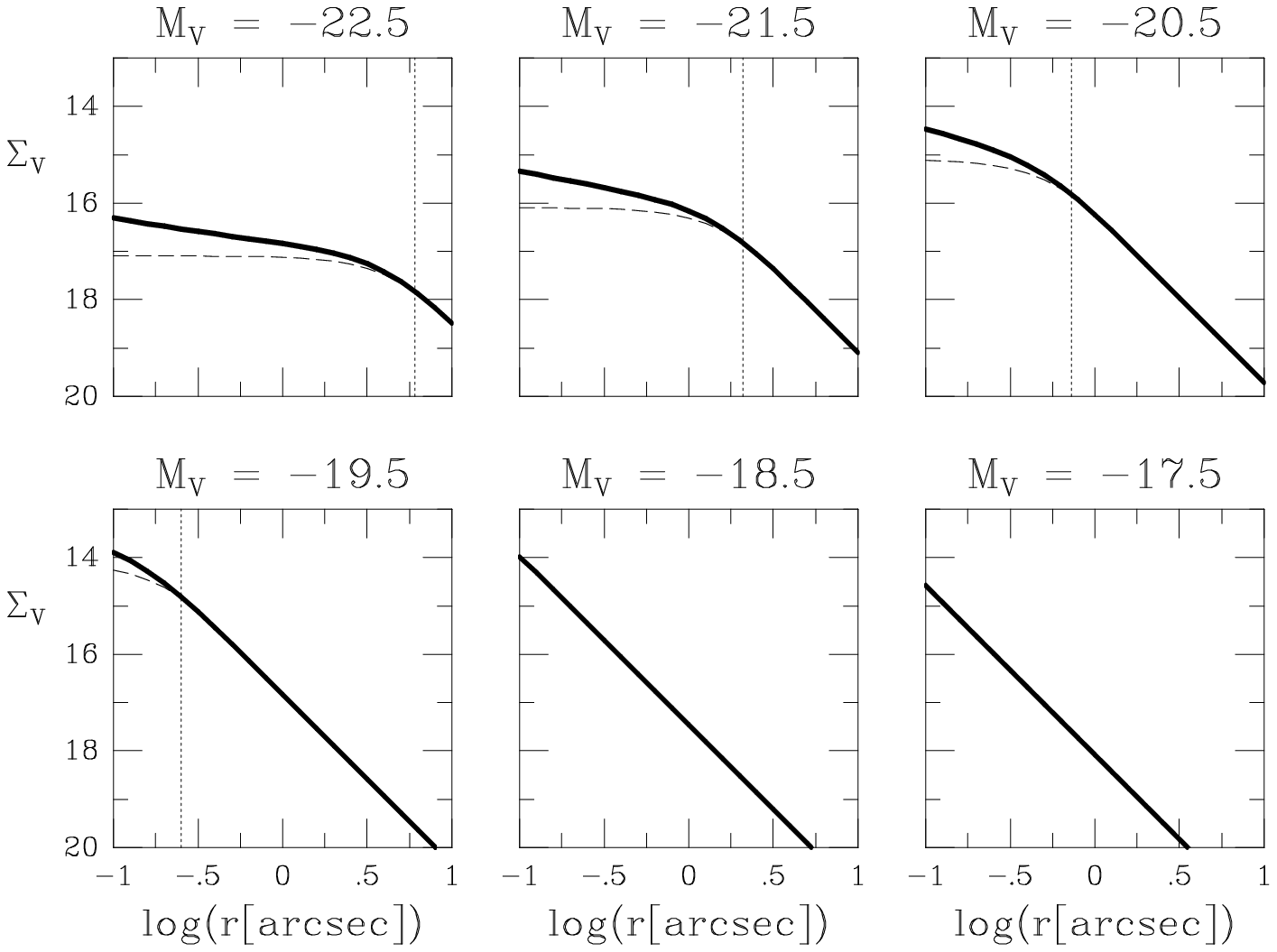}}
\ifsubmode
\vskip3.0truecm
\addtocounter{figure}{1}
\centerline{Figure~\thefigure}
\else\figcaption{\figcapvirgoprof}\fi
\end{figure}


\clearpage
\begin{figure}
\centerline{\epsfbox{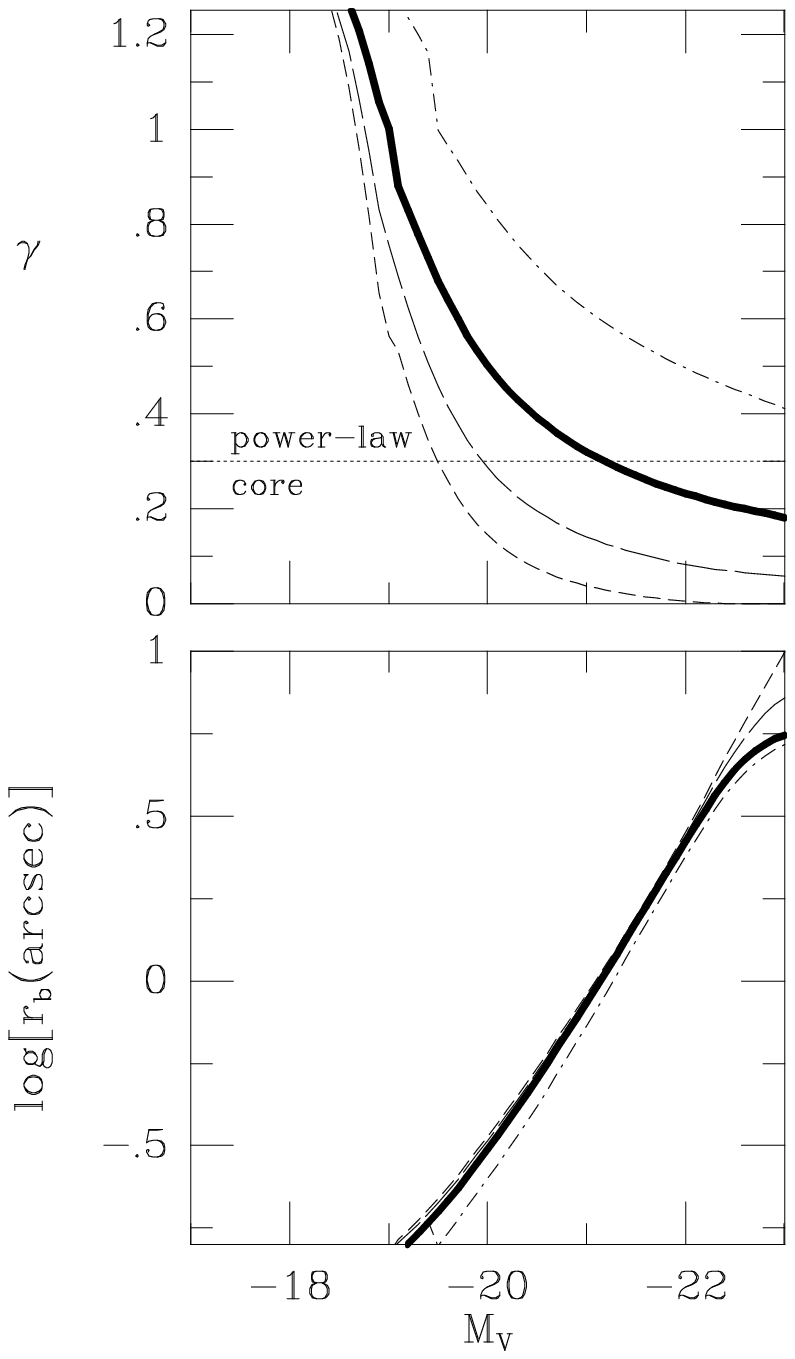}}
\ifsubmode
\vskip3.0truecm
\addtocounter{figure}{1}
\centerline{Figure~\thefigure}
\else\figcaption{\figcapvirgonuk}\fi
\end{figure}


\ifsubmode\else
\clearpage
\begin{figure}
\centerline{({\tt Figure on next two pages})}
\vskip3.0truecm
\figcaption{\figcapindigalsC}
\addtocounter{figure}{-1}
\end{figure}
\fi


\clearpage
\begin{figure}
\centerline{\epsfbox{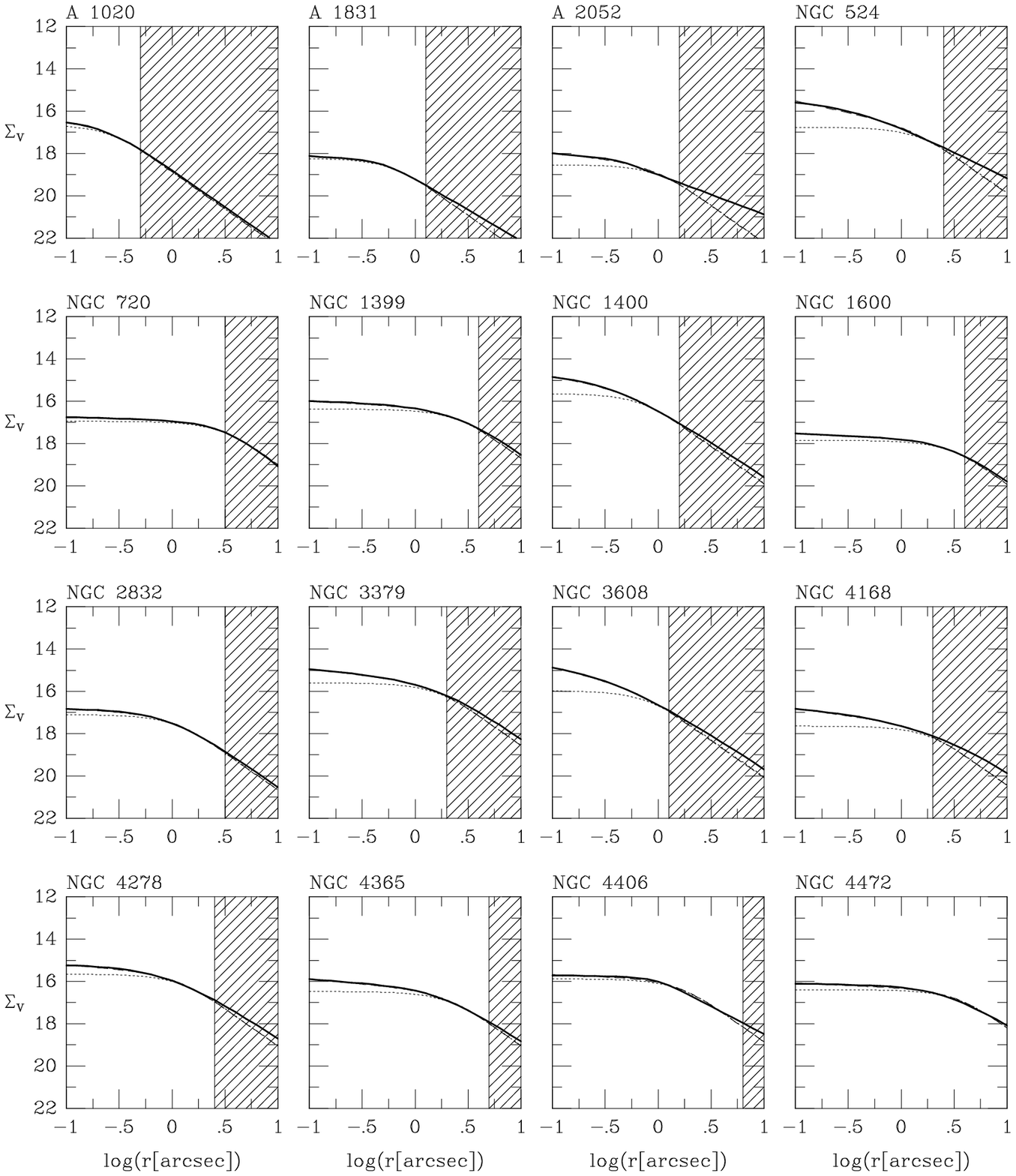}}
\ifsubmode
\vskip1.0truecm
\addtocounter{figure}{1}
\centerline{Figure~\thefigure (first part)}
\else\figcaption{(first part).}
\addtocounter{figure}{-1}\fi
\end{figure}


\clearpage
\begin{figure}
\centerline{\epsfbox{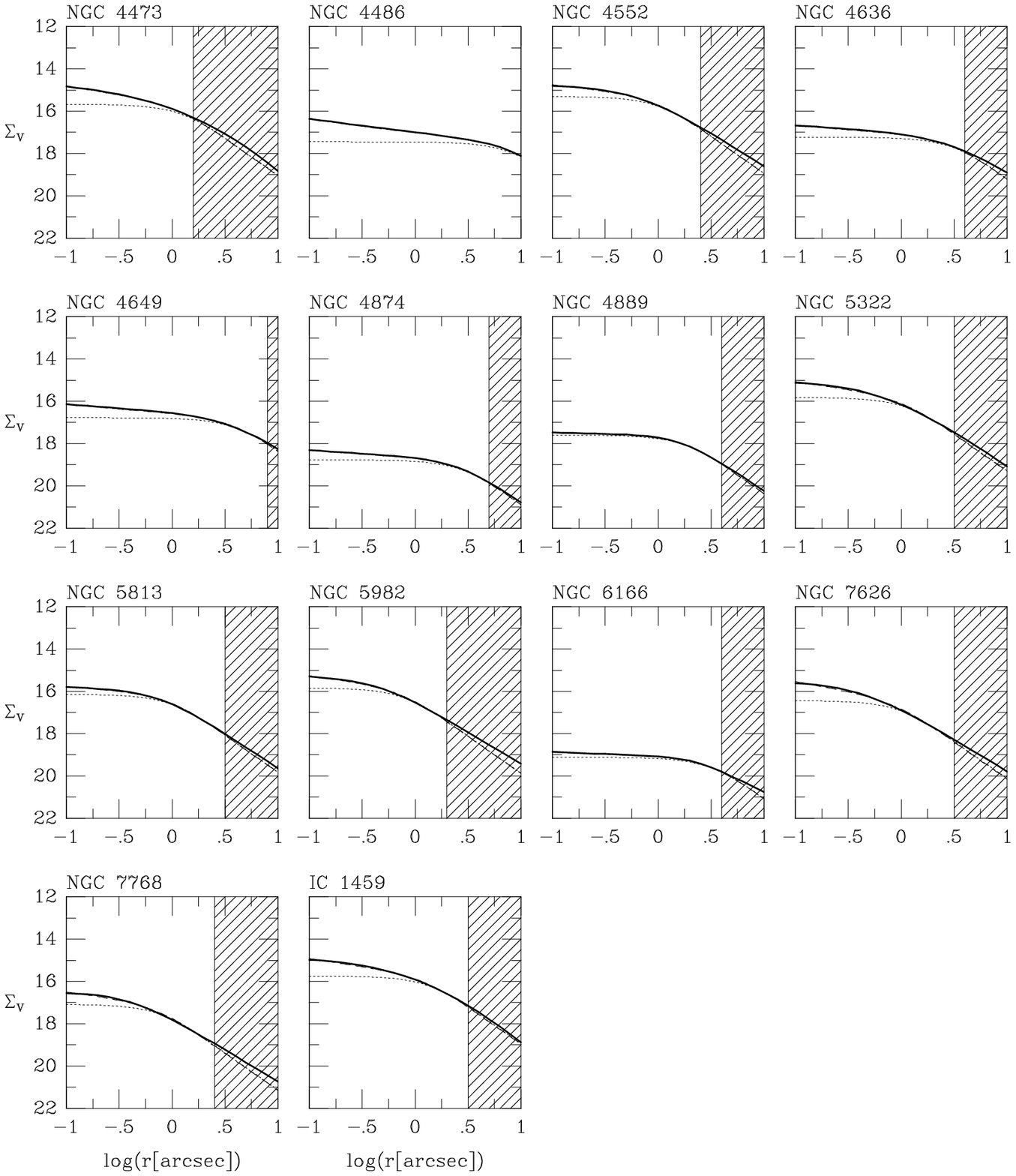}}
\ifsubmode
\vskip1.0truecm
\addtocounter{figure}{0}
\centerline{Figure~\thefigure (second part)}
\else\figcaption{(second part).}
\fi
\end{figure}


\clearpage
\begin{figure}
\centerline{\epsfbox{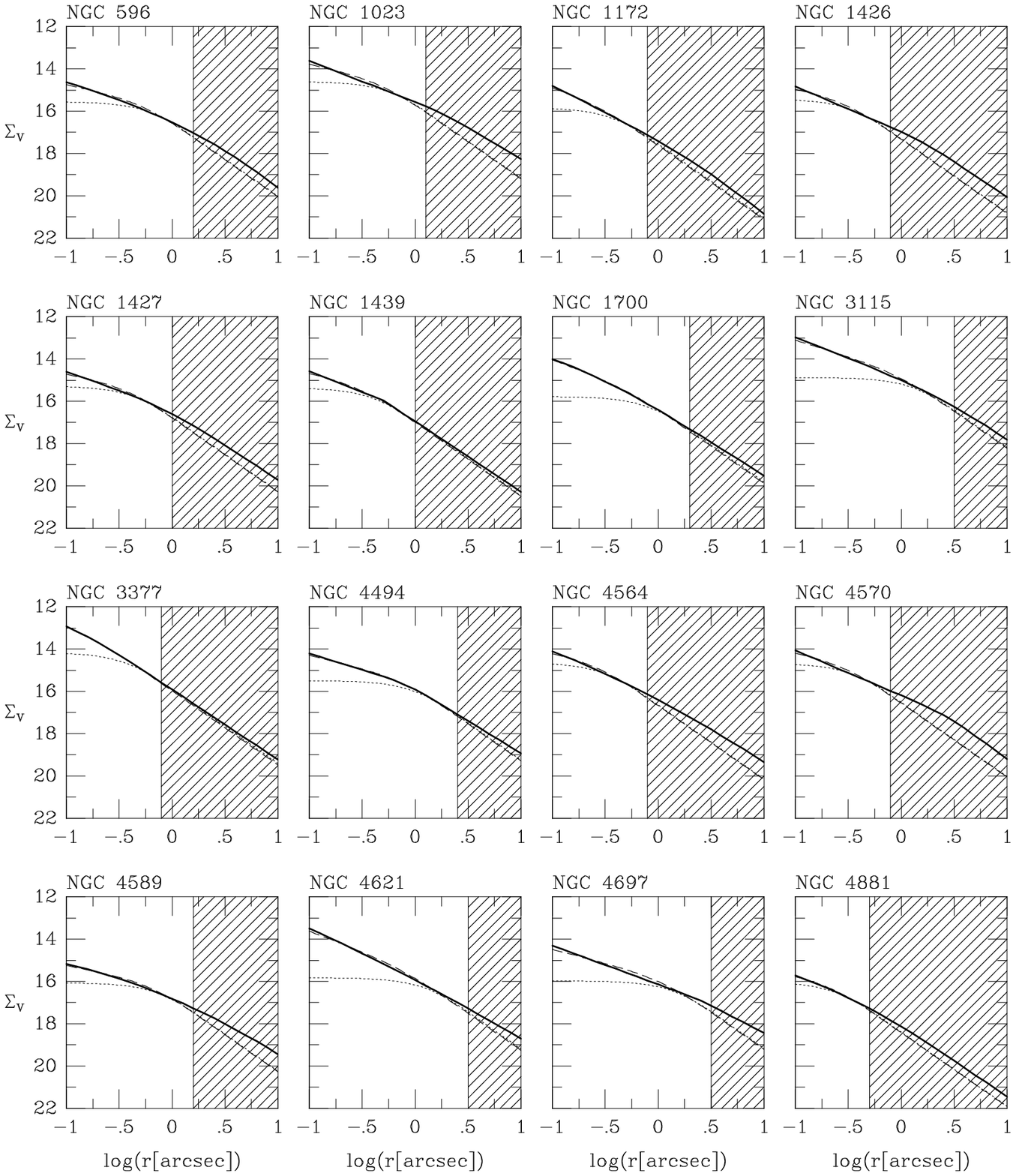}}
\ifsubmode
\vskip1.0truecm
\addtocounter{figure}{1}
\centerline{Figure~\thefigure}
\else\figcaption{\figcapindigalsPL}\fi
\end{figure}


\clearpage
\begin{figure}
\centerline{\epsfbox{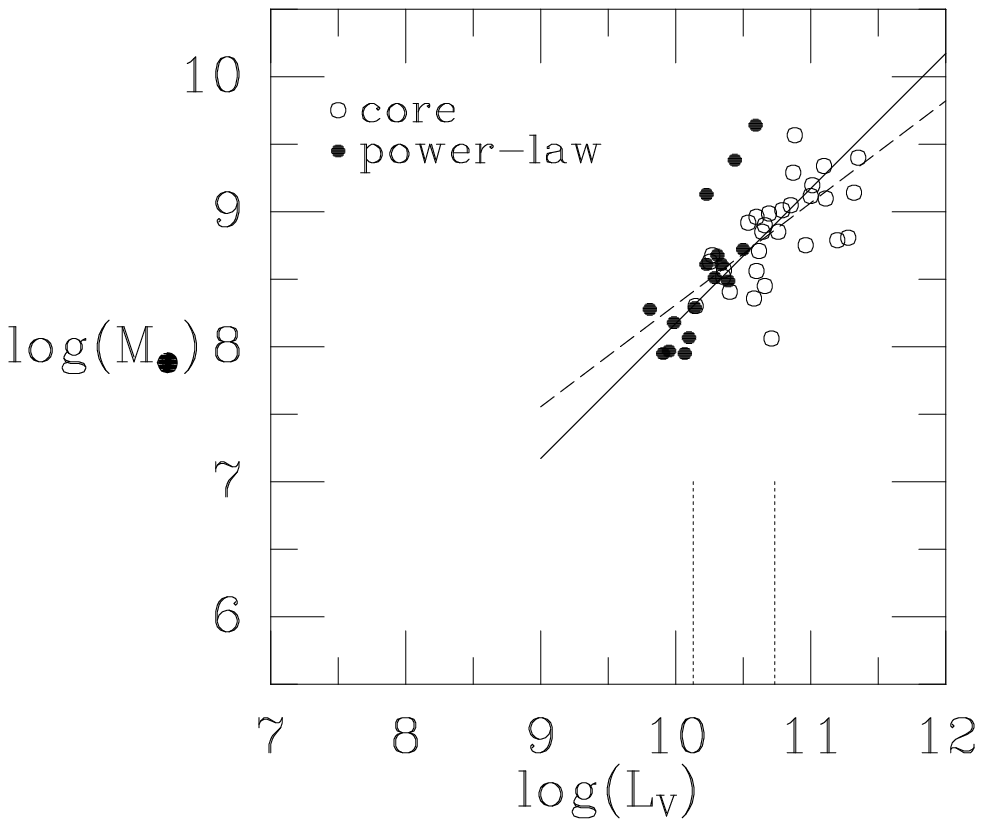}\qquad
            \epsfbox{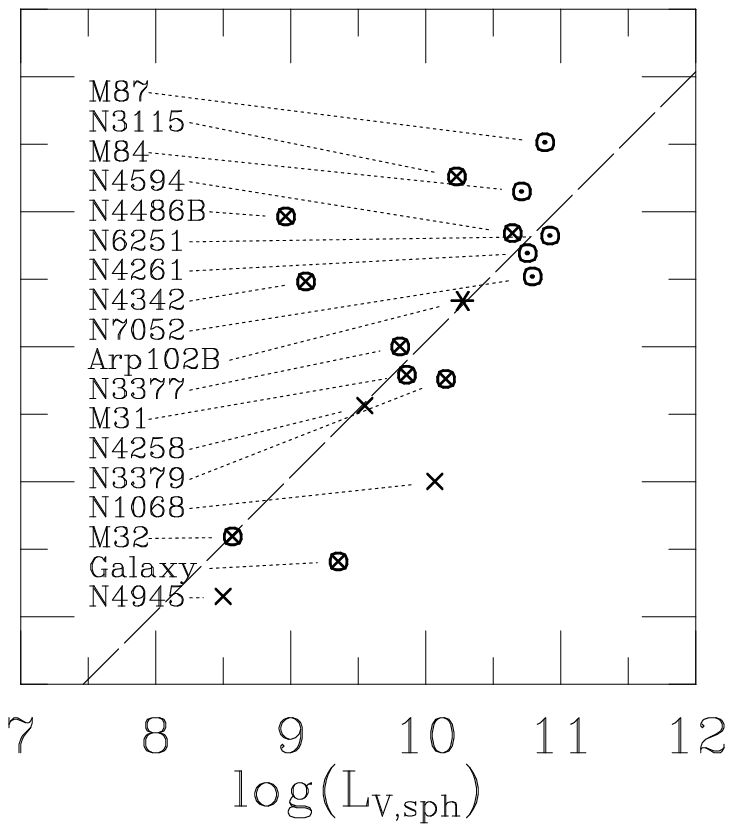}}
\ifsubmode
\vskip3.0truecm
\addtocounter{figure}{1}
\centerline{Figure~\thefigure}
\else\figcaption{\figcapBHmassdist}\fi
\end{figure}


\clearpage
\begin{figure}
\epsfxsize=16.0truecm
\centerline{\epsfbox{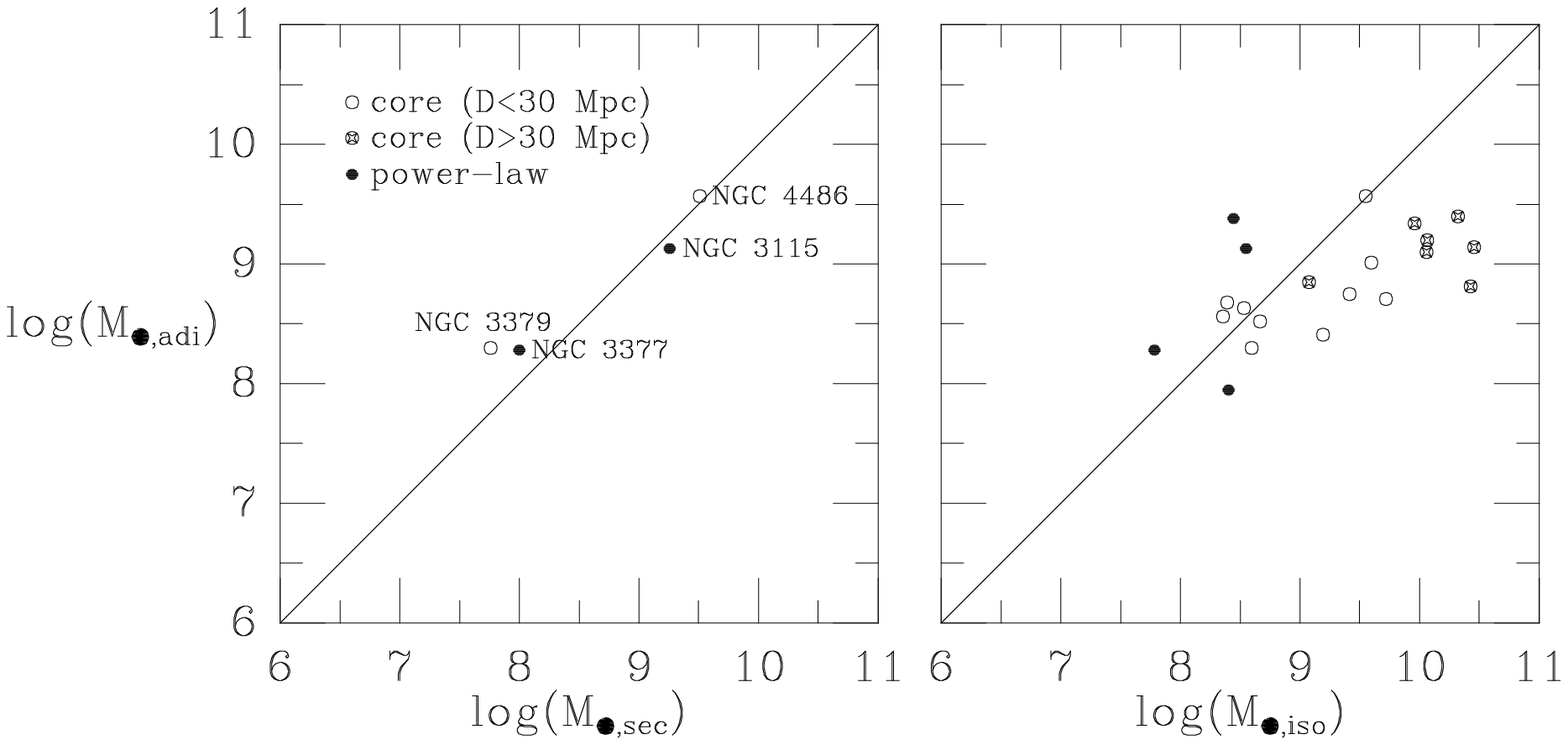}}
\ifsubmode
\vskip3.0truecm
\addtocounter{figure}{1}
\centerline{Figure~\thefigure}
\else\figcaption{\figcapBHcompare}\fi
\end{figure}


\clearpage
\begin{figure}
\centerline{\epsfbox{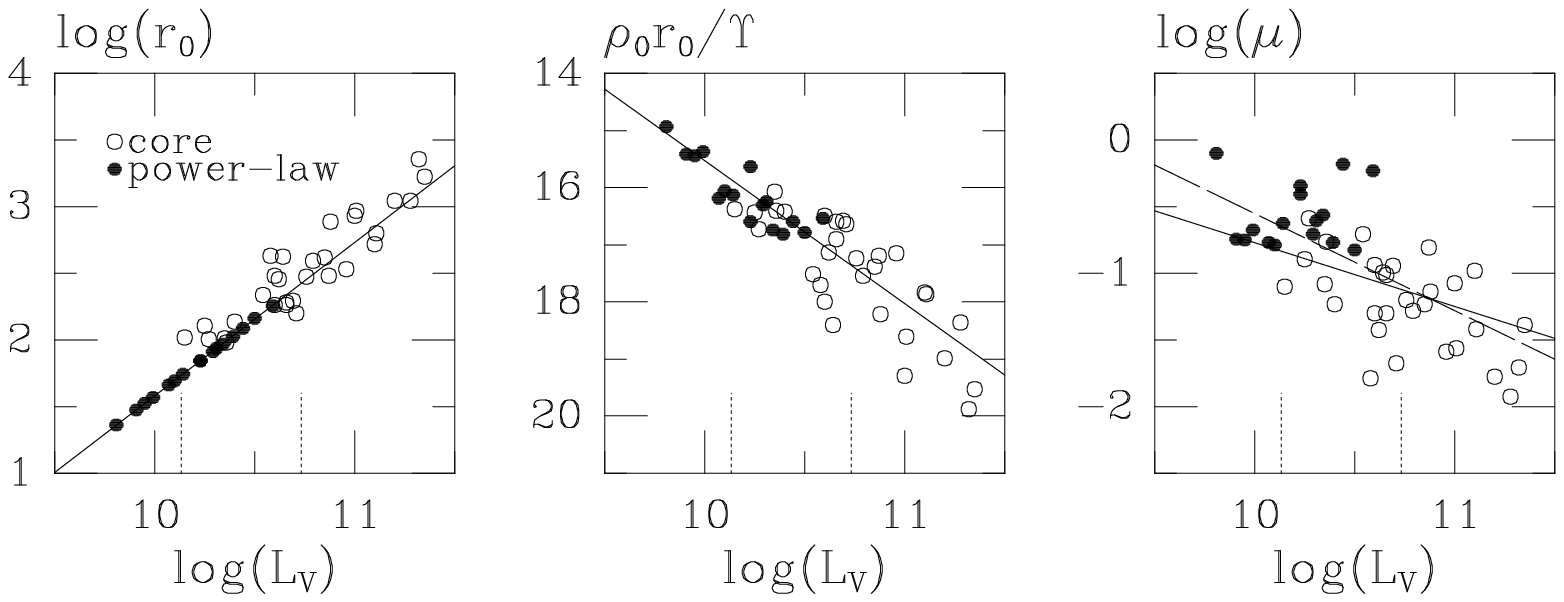}}
\ifsubmode
\vskip3.0truecm
\addtocounter{figure}{1}
\centerline{Figure~\thefigure}
\else\figcaption{\figcapmodpar}\fi
\end{figure}


\clearpage
\begin{figure}
\centerline{\epsfbox{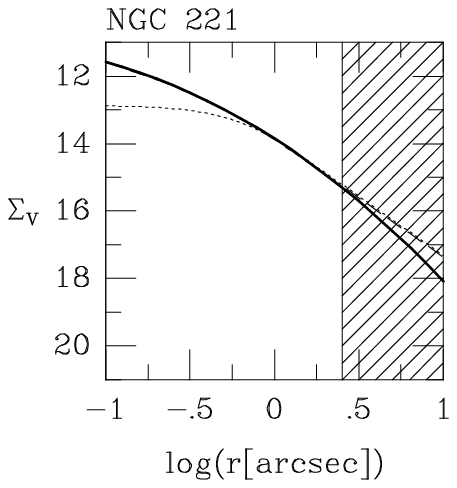}}
\ifsubmode
\vskip3.0truecm
\addtocounter{figure}{1}
\centerline{Figure~\thefigure}
\else\figcaption{\figcapindigalsM}\fi
\end{figure}


\fi


\clearpage
\ifsubmode\pagestyle{empty}\fi


\begin{deluxetable}{lccccccccc@{\extracolsep{-2pt}}c}
\scriptsize
\tablecaption{Sample properties and model results for core 
galaxies\label{t:sampleC}}
\tablehead{
\colhead{name} & \colhead{source} & \colhead{$D$} & \colhead{$\MV$} & 
\colhead{$\log L$} & \colhead{$r_0$} & \colhead{$\rho_0 r_0 /\Upsilon$} &
\colhead{$\log\mu$} & \colhead{$\log(\Mbh/\Upsilon)$} & 
\colhead{$\log\Mbh$} & \colhead{RMS} \\
 & & \colhead{(Mpc)} & & \colhead{($\Lsun$)} & \colhead{($''$)} & 
\colhead{(mag/$['']^2$)} & & \colhead{($\Lsun$)} & \colhead{($\Msun$)} &
\colhead{(mag/$['']^2$)} \\
\colhead{(1)} & \colhead{(2)} & \colhead{(3)} &
\colhead{(4)} & \colhead{(5)} & \colhead{(6)} &
\colhead{(7)} & \colhead{(8)} & \colhead{(9)} &
\colhead{(10)} & \colhead{(11)} \\
}
\ifsubmode\renewcommand{\arraystretch}{0.68}\fi
\startdata
A 1020    & F97 & 243.8 & $-22.29$ & 10.85 &  0.35 & 17.39 & $-1.23$ & 8.22 & 9.05 & 0.00 \\
A 1831    & F97 & 280.9 & $-23.16$ & 11.20 &  0.81 & 18.99 & $-1.77$ & 7.90 & 8.79 & 0.01 \\
A 2052    & F97 & 132.0 & $-22.66$ & 11.00 &  1.34 & 19.29 & $-1.08$ & 8.26 & 9.12 & 0.04 \\
NGC 524   & F97 &  23.1 & $-21.51$ & 10.54 &  1.95 & 17.52 & $-0.71$ & 8.15 & 8.92 & 0.05 \\
NGC 720   & F97 &  22.6 & $-21.62$ & 10.58 &  3.91 & 17.71 & $-1.78$ & 7.58 & 8.36 & 0.00 \\
NGC 1399  & F97 &  17.9 & $-21.71$ & 10.62 &  3.29 & 17.13 & $-1.42$ & 7.82 & 8.71 & 0.02 \\
NGC 1400  & F97 &  21.5 & $-21.06$ & 10.36 &  0.92 & 16.40 & $-0.76$ & 7.83 & 8.57 & 0.03 \\
NGC 1600  & F97 &  50.2 & $-22.70$ & 11.01 &  3.85 & 18.61 & $-1.56$ & 8.12 & 9.20 & 0.01 \\
NGC 2832  & F97 &  90.2 & $-22.95$ & 11.11 &  1.45 & 17.87 & $-1.42$ & 8.22 & 9.10 & 0.03 \\
NGC 3379  & F97 &   9.9 & $-20.55$ & 10.15 &  2.17 & 16.37 & $-1.10$ & 7.57 & 8.30 & 0.01 \\
NGC 3608  & F97 &  20.3 & $-20.84$ & 10.27 &  1.03 & 16.73 & $-0.59$ & 7.91 & 8.68 & 0.02 \\
NGC 4168  & F97 &  36.4 & $-21.76$ & 10.64 &  2.40 & 18.40 & $-0.99$ & 8.08 & 8.85 & 0.02 \\
NGC 4278  & C97 &  17.5 & $-21.16$ & 10.40 &  1.62 & 16.42 & $-1.23$ & 7.66 & 8.41 & 0.03 \\
NGC 4365  & F97 &  22.0 & $-22.06$ & 10.76 &  2.80 & 17.24 & $-1.20$ & 8.04 & 8.85 & 0.03 \\
NGC 4406  & C97 &  15.3 & $-21.94$ & 10.71 &  2.13 & 16.64 & $-1.67$ & 7.25 & 8.06 & 0.07 \\
NGC 4472  & F97 &  15.3 & $-22.57$ & 10.96 &  4.57 & 17.15 & $-1.59$ & 7.80 & 8.75 & 0.05 \\
NGC 4473  & B96 &  15.8 & $-20.80$ & 10.25 &  1.66 & 16.43 & $-0.89$ & 7.92 & 8.63 & 0.02 \\
NGC 4486  & F97 &  15.3 & $-22.38$ & 10.88 & 10.35 & 18.21 & $-1.14$ & 8.53 & 9.57 & 0.02 \\
NGC 4552  & F97 &  15.3 & $-21.05$ & 10.35 &  1.38 & 16.07 & $-1.08$ & 7.69 & 8.52 & 0.04 \\
NGC 4636  & F97 &  15.3 & $-21.67$ & 10.60 &  4.09 & 17.99 & $-1.30$ & 7.65 & 8.56 & 0.02 \\
NGC 4649  & F97 &  15.3 & $-22.14$ & 10.79 &  5.32 & 17.54 & $-1.28$ & 8.08 & 9.01 & 0.02 \\
NGC 4874  & F97 &  93.3 & $-23.54$ & 11.35 &  3.73 & 19.53 & $-1.39$ & 8.44 & 9.40 & 0.01 \\
NGC 4889  & F97 &  93.3 & $-23.36$ & 11.28 &  2.45 & 18.36 & $-1.93$ & 8.00 & 8.81 & 0.01 \\
NGC 5322  & C97 &  26.3 & $-21.90$ & 10.69 &  1.54 & 16.58 & $-0.94$ & 8.19 & 8.99 & 0.04 \\
NGC 5813  & F97 &  28.3 & $-21.81$ & 10.66 &  1.33 & 16.90 & $-1.30$ & 7.65 & 8.45 & 0.03 \\
NGC 5982  & C97 &  37.4 & $-21.83$ & 10.66 &  1.06 & 16.59 & $-1.01$ & 8.10 & 8.90 & 0.03 \\
NGC 6166  & F97 & 112.5 & $-23.47$ & 11.32 &  4.15 & 19.88 & $-1.71$ & 8.24 & 9.14 & 0.01 \\
NGC 7626  & C97 &  46.4 & $-22.34$ & 10.87 &  1.34 & 17.20 & $-0.81$ & 8.45 & 9.29 & 0.05 \\
NGC 7768  & F97 & 103.1 & $-22.93$ & 11.10 &  1.05 & 17.84 & $-0.98$ & 8.50 & 9.34 & 0.05 \\
IC 1459   & C97 &  20.7 & $-21.68$ & 10.60 &  1.83 & 16.50 & $-0.94$ & 8.17 & 8.96 & 0.04 \\
\enddata
\tablecomments{\scriptsize
Column~(1) is the galaxy name. `A' objects are first-brightest Abell
cluster galaxies. Column~(2) indicates the source of the nuker-law
parameters used in this paper; Faber \etal (1997) [F97], Carollo \etal
(1997) [C97] or Byun \etal (1996) [B96]. Columns~(3)--(5) list the
distance, absolute {\VV}-band magnitude and total {\VV}-band
luminosity, respectively, generally taken from the source in
column~(2). Columns~(6)--(8) list the parameters of the best-fitting
adiabatic BH growth model: the core radius $r_0$ and intensity scale
$\rho_0 r_0 / \Upsilon$ of the initial isothermal model, both in
observed units, and the dimensionless BH mass $\mu$. Columns~(9)
and~(10) list the implied $\Mbh/\Upsilon$ and $\Mbh$ in physical
units. The {\VV}-band mass-to-light ratio $\Upsilon$ was taken from
Magorrian \etal (1998), where available, and otherwise from
equation~(\ref{Upscor}). Column~(11) lists the RMS residual of the
adiabatic BH growth model fit to the observed nuker-law
parameterization, over the non-hatched region in
Figure~\ref{f:indigalsC} or~\ref{f:indigalsPL}. All quantities assume
$H_0 = 80 \kms \Mpc^{-1}$.}
\end{deluxetable}


\begin{deluxetable}{lcccccccccc}
\scriptsize
\tablecaption{Sample properties and model results for power-law
galaxies\label{t:samplePL}}
\tablehead{
\colhead{name} & \colhead{source} & \colhead{$D$} & \colhead{$\MV$} & 
\colhead{$\log L$} & \colhead{$r_0$} & \colhead{$\rho_0 r_0 /\Upsilon$} &
\colhead{$\log\mu$} & \colhead{$\log(\Mbh/\Upsilon)$} & 
\colhead{$\log\Mbh$} & \colhead{RMS} \\
 & & \colhead{(Mpc)} & & \colhead{($\Lsun$)} & \colhead{($''$)} & 
\colhead{(mag/$['']^2$)} & & \colhead{($\Lsun$)} & \colhead{($\Msun$)} &
\colhead{(mag/$['']^2$)} \\
\colhead{(1)} & \colhead{(2)} & \colhead{(3)} &
\colhead{(4)} & \colhead{(5)} & \colhead{(6)} &
\colhead{(7)} & \colhead{(8)} & \colhead{(9)} &
\colhead{(10)} & \colhead{(11)} \\
}
\ifsubmode\renewcommand{\arraystretch}{0.68}\fi
\startdata
NGC 596   & F97 &  21.2 & $-20.90$ & 10.29 &  0.80 & 16.31 & $-0.71$ & 7.78 & 8.51 & 0.10 \\
NGC 1023  & F97 &  10.2 & $-20.14$ &  9.99 &  0.75 & 15.37 & $-0.67$ & 7.50 & 8.18 & 0.13 \\
NGC 1172  & F97 &  29.8 & $-20.74$ & 10.23 &  0.48 & 16.60 & $-0.34$ & 7.89 & 8.61 & 0.07 \\
NGC 1426  & F97 &  21.5 & $-20.35$ & 10.07 &  0.44 & 16.19 & $-0.77$ & 7.26 & 7.95 & 0.11 \\
NGC 1427  & C97 &  17.9 & $-20.43$ & 10.10 &  0.57 & 16.05 & $-0.79$ & 7.37 & 8.07 & 0.09 \\
NGC 1439  & C97 &  21.5 & $-20.51$ & 10.14 &  0.53 & 16.13 & $-0.63$ & 7.59 & 8.29 & 0.05 \\
NGC 1700  & F97 &  35.5 & $-21.65$ & 10.59 &  1.05 & 16.54 & $-0.23$ & 8.85 & 9.64 & 0.04 \\
NGC 3115  & F97 &   8.4 & $-20.75$ & 10.23 &  1.72 & 15.64 & $-0.40$ & 8.21 & 9.13 & 0.10 \\
NGC 3377  & F97 &   9.9 & $-19.70$ &  9.81 &  0.48 & 14.93 & $-0.10$ & 7.83 & 8.28 & 0.02 \\
NGC 4494  & C97 &  14.0 & $-20.94$ & 10.31 &  1.27 & 16.25 & $-0.61$ & 7.95 & 8.68 & 0.04 \\
NGC 4564  & F97 &  15.3 & $-19.94$ &  9.91 &  0.40 & 15.41 & $-0.75$ & 7.23 & 7.95 & 0.09 \\
NGC 4570  & F97 &  15.3 & $-20.04$ &  9.95 &  0.45 & 15.45 & $-0.75$ & 7.30 & 7.97 & 0.11 \\
NGC 4589  & C97 &  22.9 & $-21.14$ & 10.39 &  0.96 & 16.82 & $-0.77$ & 7.74 & 8.49 & 0.09 \\
NGC 4621  & F97 &  15.3 & $-21.27$ & 10.44 &  1.64 & 16.59 & $-0.18$ & 8.54 & 9.38 & 0.10 \\
NGC 4697  & F97 &  10.5 & $-21.03$ & 10.34 &  1.84 & 16.74 & $-0.56$ & 7.87 & 8.61 & 0.15 \\
NGC 4881  & B96 &  93.3 & $-21.41$ & 10.50 &  0.32 & 16.79 & $-0.83$ & 7.95 & 8.72 & 0.06 \\
\enddata
\tablecomments{\scriptsize
See notes to Table~\ref{t:sampleC} for descriptions of the table
entries. The progenitor core radius $r_0$ for power-law galaxies was
not obtained from a fit to the data, but was fixed a priori to the
value predicted by equation~(\ref{rrhoisorel}), as motivated in the
text.}
\end{deluxetable}


\begin{deluxetable}{lcccccccccc}
\scriptsize
\tablecaption{Properties and model results for M32\label{t:sampleM}}
\tablehead{
\colhead{name} & \colhead{source} & \colhead{$D$} & \colhead{$\MV$} & 
\colhead{$\log L$} & \colhead{$r_0$} & \colhead{$\rho_0 r_0 /\Upsilon$} &
\colhead{$\log\mu$} & \colhead{$\log(\Mbh/\Upsilon)$} & 
\colhead{$\log\Mbh$} & \colhead{RMS} \\
 & & \colhead{(Mpc)} & & \colhead{($\Lsun$)} & \colhead{($''$)} & 
\colhead{(mag/$['']^2$)} & & \colhead{($\Lsun$)} & \colhead{($\Msun$)} &
\colhead{(mag/$['']^2$)} \\
\colhead{(1)} & \colhead{(2)} & \colhead{(3)} &
\colhead{(4)} & \colhead{(5)} & \colhead{(6)} &
\colhead{(7)} & \colhead{(8)} & \colhead{(9)} &
\colhead{(10)} & \colhead{(11)} \\
}
\ifsubmode\renewcommand{\arraystretch}{0.68}\fi
\startdata
NGC 221   & - &   0.8 & $-16.60$ &  8.57 &  0.81 & 13.63 & $-0.38$ & 6.34 & 6.68 & 0.02 \\
\enddata
\tablecomments{\scriptsize
See notes to Table~\ref{t:sampleC} for descriptions of the table
entries. The dash in column~(2) indicates that the nuker-law
parameters for M32 were not taken from a published source, but were
obtained from a new nuker-law fit to the data of Lauer \etal (1992b).
The parameters of this fit are: $r_b = 1.58''$, $I_b = 14.77$
mag/arcsec$^2$, $\alpha = 0.50$, $\beta = 2.89$ and $\gamma = 0.00$.
The progenitor core radius $r_0$ of the adiabatic BH growth model in
column~(6) was not fixed a priori (by contrast to the approach adopted
for other power-law galaxies) but was obtained from a fit to the
data.}
\end{deluxetable}



\end{document}